\documentclass[twocolumn,trackchanges]{aastex631}

\usepackage{graphicx}
\usepackage{subfigure}
\usepackage{paralist}
\usepackage{amssymb}
\usepackage{bm}
\usepackage{bbding}
\usepackage[fleqn]{amsmath}
\usepackage{multirow}
\hypersetup{linkcolor=red,citecolor=blue,filecolor=cyan,urlcolor=magenta}

\submitjournal{AJ}

\shorttitle{Spin-orbit coupling}
\shortauthors{Lei}

\begin{document}

\title{Spin-orbit coupling of the ellipsoidal secondary in a binary asteroid system}

\correspondingauthor{Hanlun Lei}
\email{leihl@nju.edu.cn}

\author{Hanlun Lei}
\affiliation{School of Astronomy and Space Science, Nanjing University, Nanjing 210023, China}
\affiliation{Key Laboratory of Modern Astronomy and Astrophysics in Ministry of Education, Nanjing University, Nanjing 210023, China}



\begin{abstract}
In our Solar system, spin-orbit coupling is a common phenomenon in binary asteroid systems, where the mutual orbits are no longer invariant due to exchange of angular momentum between translation and rotation. In this work, dynamical structures in phase space are explored for the problem of spin-orbit coupling by taking advantage of analytical and numerical methods. In particular, the technique of Poincar\'e sections is adopted to reveal numerical structures, which are dependent on the total angular momentum, the Hamiltonian, mass ratio and asphericity parameter. Analytical study based on perturbative treatments shows that high-order and/or secondary spin-orbit resonances are responsible for numerical structures arising in Poincar\'e sections. Analytical solutions are applied to (65803) Didymos, (80218) ${\rm VO}_{123}$ and (4383) Suruga to reveal their phase-space structures, showing that there is a high possibility for them to locate inside secondary 1:1 spin-orbit resonance.
\end{abstract}



\section{Introduction}
\label{Sect1}

Spin-orbit resonance takes place if there is a commensurability between the rotational period of one object and its orbital period moving around the central object \citep{Goldreich1966Spin,peale1977rotation,murray1999solar}. In our Solar system, it is a common phenomenon under Sun-planet \citep{peale1965rotation,lemaitre20063}, planet-satellite \citep{peale1969generalized} and binary asteroid systems \citep{scheeres2006dynamical,naidu2015near,pravec2016binary,pravec2019asteroid}. In particular, chaotic rotations have been observed for Saturn's satellite Hyperion \citep{wisdom1984chaotic} as well as Pluto's satellites Nyx and Hydra \citep{correia2015spin,showalter2015resonant}. 

In the conventional model of spin-orbit resonances, the total angular momentum is dominated by the orbital angular momentum and exchange of angular momentum can be ignored, thus it is approximated that their mutual orbit remains invariant \citep{Goldreich1966Spin,Celletti1990AnalysisI,Celletti1990AnalysisII}. It is a good model to describe spin-orbit resonances under Sun-planet and planet-satellite systems \citep{murray1999solar}. Under the classical model, there are a series of spin-orbit resonances, depending on the asphericity parameter $\alpha$ and eccentricity of mutual orbit \citep{flynn2005second,nadoushan2016geography}. Among them, the synchronous spin-orbit resonance is the strongest one. In particular, the 1:1 resonance is zero order in eccentricity, 3:2 and 1:2 resonances are first order and 4:2 (2:1) are second order in eccentricity, and so on. According to the criterion proposed in \citet{chirikov1979universal}, overlap of nearby spin-orbit resonances may lead to chaos. Based on this theory, the onset mechanism of chaos was discussed by \citet{wisdom1984chaotic} in order to understand chaotic rotations of Saturn's satellite Hyperion. Such an overlap criterion of chaos is updated by \citet{jafari2015widespread} and \citet{jafari2016chirikov}. Spin-orbit resonances can also be explored from the viewpoint of periodic orbits \citep{celletti2000hamiltonian}. Considering quadrupole–quadrupole interactions with circular mutual orbits, \citet{batygin2015spin} studied spin-spin coupling of binary objects in the Solar system, suggesting that spin–spin coupling may be consequential for highly elongated, tightly orbiting binary objects. Under the assumption of invariant orbit, \citet{nadoushan2016geography} investigated a geography of spin-orbit-spin resonances by considering the coupling terms arising in the fourth-order Hamiltonian model. If the frequency of small-amplitude oscillation inside the primary resonance and the orbital frequency are commensurable, secondary spin-orbit resonances may happen. It is known that Enceladus is possible to be located inside the secondary 3:1 spin-orbit resonance \citep{wisdom2004spin}. A series of secondary spin-orbit resonances (including 1:1, 2:1 and 3:1) are analytically studied in \citet{gkolias2016theory} and \citet{gkolias2019accurate} by taking advantage of Lie-series transformation theory. Recently, \citet{lei2024dynamical} systematically explored global structures in the $(\dot\theta,\alpha)$ space and the author concluded that the V-shape structure is sculpted by the synchronous primary resonance, those minute structures inside the V-shape region are dominated by secondary resonances and those structures outside the V-shape region are governed by high-order resonances.    

When the rotational angular momentum is not negligible compared to the orbital angular momentum, the conventional model of spin-orbit resonance cannot catch dynamical behaviors of binary systems. In this circumstance, the orbit and rotation of a binary system are coupled in evolution and thus it becomes a spin-orbit coupling problem, corresponding to the full two-body problem, where the mutual potential, force and torque are formulated by \citet{hou2017mutual}. Due to the coupling, there exists an exchange of angular momentum between translation and rotation, thus the mutual orbit is no longer invariant. Considering spin-orbit coupling, \citet{naidu2015near} made an examination about the rotational regimes of near-Earth binary steroids by using surfaces of section. Under the planar ellipsoid-ellipsoid configuration, the spin-orbit, spin-spin and spin-orbit-spin resonances are analytically studied by \citet{hou2017note}, showing that the libration center of spin-orbit resonances may vary with the mass ratio and mutual distance. Under the planar sphere-ellipsoid configuration, the 1:1 spin-orbit resonance and its dynamical stability are studied by taking advantage of the approach of periodic orbits in \citet{wang2020secondary} and \citet{wang2022stability}. Extending to three-dimensional configurations, \citet{tan2023attitude} performed a systematic stability analysis about synchronous binary asteroids under the sphere–ellipsoid model. Besides spin-orbit coupling, thermal effects may change rotation states of binary objects in the Solar system \citep{scheeres2007dynamical,cuk2005effects,mcmahon2010secular,fang2011near}.  

Regarding the strength of spin-orbit coupling, \citet{jafari2023surfing} provided a criterion of dynamical closeness: if the ratio of rotation-to-orbit angular momentum of a binary system is greater than 10 percent, it is a close binary system. In addition, they explored phase-space structures by taking advantage of the fast Lyapunov index (FLI). In particular, substructures inside and/or outside the region of synchronous spin-orbit resonance are visible from their dynamical maps. However, dynamical explanations and emergence condition about these substructures in dynamical maps are not clear. The purpose of this work is to address this problem. Firstly, dynamical structures are revealed by means of plotting Poincar\'e sections and their dependence on the total angular momentum, the magnitude of Hamiltonian, ratio of secondary-to-primary mass and asphericity parameter are discussed. Secondly, dynamical structures arising in Poicnar\'e sections are analytically investigated following the similar approach adopted in \citet{lei2024dynamical}. Results show that dynamical structures arising in Poincar\'e sections (or dynamical maps) are triggered by high-order and/or secondary spin-orbit resonances. At last, analytical solutions are applied to three binary asteroid systems, including (65803) Didymos, (80218) ${\rm VO}_{123}$ and (4383) Suruga, in order to reveal their dynamical structures.

The remaining part of this work is organized as follows. In Section \ref{Sect2}, Hamiltonian model is formulated for describing spin-orbit coupling and two parameters are introduced to characterize the total angular momentum and Hamiltonian. Numerical explorations based on Poincar\'e surfaces of sections are performed in Section \ref{Sect3}. Analytical study based on elliptic expansion of Hamiltonian and perturbative treatments are made in Section \ref{Sect4} to study high-order and secondary spin-orbit resonances and applications to three binary asteroid systems are shown in Section \ref{Sect5}. Conclusions are summarized in Section \ref{Sect6}.

\begin{figure}
\centering
\includegraphics[width=0.8\columnwidth]{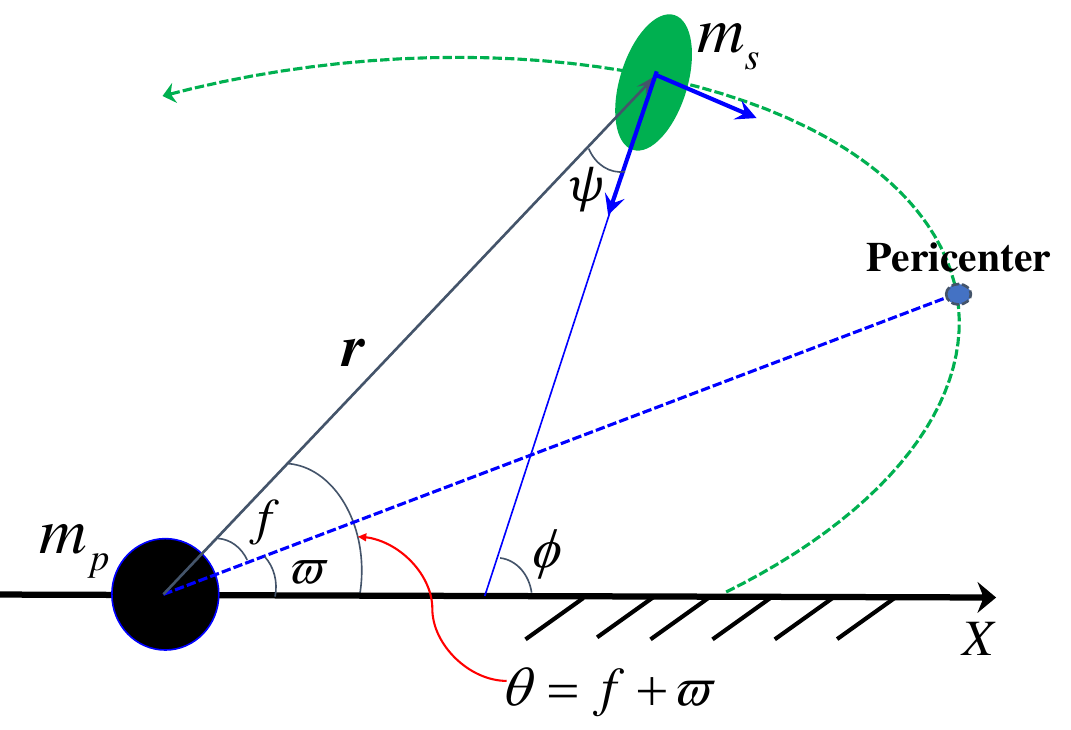}
\caption{Schematic diagram of variables, used for describing translational and rotational motions, in the binary asteroid system composed of a spherical primary with mass $m_p$ and an ellipsoidal secondary with mass $m_s$. The second-to-primary mass ratio is denoted as $\alpha_m = m_s/m_p <1$. In the planar problem, the $X$-axis directs towards an arbitrary fixed direction. The position vector from the primary towards the secondary is denoted by $\bm r$, the true longitude of the secondary is denoted by $\theta=f+\varpi$ where $\varpi$ is the longitude of pericenter and $f$ is the true anomaly, and the rotational angle of the secondary relative to the $X$-axis of an inertial frame is represented by $\phi$. The relative angle is defined by $\psi = \phi-\theta = (\phi-\varpi) - f$.}
\label{Fig1}
\end{figure}

\section{Dynamical model}
\label{Sect2}

In this work, we concentrate on the dynamics of spin-orbit coupling in a binary asteroid system, which is composed of a spherical primary with mass $m_p$ and an ellipsoidal secondary with mass $m_s$. Similar configuration is adopted in \citet{hou2017note}, \citet{wang2020secondary} and \citet{jafari2023surfing} for different purposes. Figure \ref{Fig1} shows the relative geometry of the planar spin-orbit coupling problem. The radius vector from the primary to the secondary is denoted by $\bm r$, the true longitude $\theta = f + \varpi$ describes the translation and the angle $\phi$ describes the rotation of the secondary. The physical semi-major axes of the secondary are denoted by $a_s$, $b_s$ and $c_s$, satisfying $a_s \ge b_s \ge c_s$, and its moments of inertia along principal axes are given by
\begin{equation}\label{Eq1}
\left\{I_1,I_2,I_3\right\}= \frac{1}{5}m_s \left\{b_s^2 + c_s^2,a_s^2 + c_s^2,a_s^2 + b_s^2\right\}.
\end{equation}

The kinetic energy of system is the sum of the orbital and rotational kinetic energy, given by
\begin{equation}\label{Eq2}
T = \frac{1}{2}m \left( {\dot r}^2 + r^2 {\dot\theta}^2\right) + \frac{1}{2}I_3 {\dot\phi}^2,
\end{equation}
where $m = \frac{m_p m_s}{m_p + m_s}$ is the reduced mass of system. The gravitational potential between the primary and the secondary takes the following form \citep{hou2017note,jafari2023surfing}\footnote{There is an error for the sign of the term containing $C_{42}$ in \citet{jafari2023surfing}.}:
\begin{equation}\label{Eq3}
\begin{aligned}
U (r,\theta,\phi) =&  - {{\cal G} m_p m_s} \left[ {\frac{1}{r} - \frac{{a_s^2{C_{20}}}}{{2{r^3}}} + \frac{{3a_s^4{C_{40}}}}{{8{r^5}}}} \right.\\
& + {\left( {\frac{{3a_s^2{C_{22}}}}{{{r^3}}} - \frac{{15a_s^4{C_{42}}}}{{2{r^5}}}} \right)\cos \left( {2\phi  - 2\theta } \right)}\\
&+ \left. \frac{{105a_s^4{C_{44}}}}{{{r^5}}}\cos \left( {4\phi  - 4\theta } \right) \right],
\end{aligned}
\end{equation}
where $\cal G$ is the Universal gravitational constant and the spherical harmonics coefficients are defined by \citep{balmino1994gravitational}
\begin{equation*}
\begin{aligned}
{C_{20}} &= \frac{1}{{10a_s^2}}\left( {2c_s^2 - a_s^2 - b_s^2} \right),\;{C_{22}} = \frac{1}{{20a_s^2}}\left( {a_s^2 - b_s^2} \right),\\
{C_{40}} &= \frac{{15}}{7}\left( {C_{20}^2 + 2C_{22}^2} \right),\;{C_{42}} = \frac{5}{7}{C_{20}}{C_{22}},\;{C_{44}} = \frac{5}{{28}}C_{22}^2.
\end{aligned}
\end{equation*}
The rotational and translational motions of the secondary are governed by the mutual gravitational potential. Due to the spin-orbit coupling, the orbit of the secondary moving around the primary is no longer invariant. 

For convenience, the following set of canonical variables are introduced:
\begin{equation}\label{Eq4}
\begin{aligned}
&r,\quad p_r = m \dot r,\\
&\theta,\quad p_\theta = m r^2 \dot\theta,\\
&\phi,\quad p_\phi = I_3 \dot\phi.
\end{aligned}
\end{equation}
Denote the set of conjugate variables as ${\bm X} = (r,p_r,\theta,p_\theta,\phi,p_\phi)^{\rm T}$. In terms of ${\bm X}$, the Hamiltonian function, governing the translation and rotation of the secondary, can be written as
\begin{equation}\label{Eq5}
{\cal H}\left(r,p_r,\theta,p_\theta,\phi,p_\phi\right) = T \left(p_r,p_\theta,p_\phi\right) + U \left(r,\phi-\theta\right). 
\end{equation}
It is observed that $\phi$ and $\theta$ appear as a combination form ($\phi-\theta$) in the potential function. Thus, we introduce the following transformation,
\begin{equation}\label{Eq6}
\begin{aligned}
&r,\quad p_r = m \dot r,\\
&\psi=\phi-\theta,\quad p_\psi = -p_\theta,\\
&\varphi = \phi,\quad p_\varphi = p_\phi + p_\theta,
\end{aligned}
\end{equation}
which is canonical with the generating function,
\begin{equation*}
S = \phi (p_\psi + p_\varphi) - \theta p_\psi.
\end{equation*}
Let us denote the new set of conjugate variables as ${\bm Y} = (r,p_r,\psi,p_\psi,\varphi,p_\varphi)^{\rm T}$. In terms of ${\bm Y}$, the Hamiltonian function can be expressed in an explicit form,
\begin{equation}\label{Eq7}
\begin{aligned}
{\cal H} =& \frac{{p_r^2}}{{2m}} + \frac{{p_\psi ^2}}{{2m{r^2}}} + \frac{1}{{2{I_3}}}{\left( {{p_\varphi } + {p_\psi }} \right)^2}\\
 &- {\cal G}{m_p}{m_s}\left[ {\frac{1}{r} - \frac{{a_s^2{C_{20}}}}{{2{r^3}}} + \frac{{3a_s^4{C_{40}}}}{{8{r^5}}} + \left( {\frac{{3a_s^2{C_{22}}}}{{{r^3}}}} \right.} \right.\\
&\left. {\left. { - \frac{{15a_s^4{C_{42}}}}{{2{r^5}}}} \right)\cos 2\psi  + \frac{{105a_s^4{C_{44}}}}{{{r^5}}}\cos 4\psi } \right].
\end{aligned}
\end{equation}
Evidently, the angular coordinate $\varphi$ is absent from Hamiltonian (\ref{Eq7}), indicating that its conjugate moment $p_{\varphi} = p_\theta + p_\phi$ is a motion integral. In particular, $p_{\varphi}$ is the total angular momentum  $G_{\rm tot}$,
\begin{equation}\label{Eq8}
p_{\varphi} = G_{\rm tot} = p_\theta + p_\phi = m r^2\dot\theta + I_3\dot\phi.
\end{equation}
Thus, we can see that Hamiltonian (\ref{Eq7}) determines a two-degree-of-freedom dynamical model, depending on the total angular momentum. Conservation of the total angular momentum shows that the
orbital and rotational angular momentum can exchange with each other.

Hamiltonian canonical relations lead to the equations of motion \citep{morbidelli2002modern},
\begin{equation}\label{Eq9}
\begin{aligned}
\dot r &= \frac{{\partial {\cal H}}}{{\partial {p_r}}},\quad {{\dot p}_r} =  - \frac{{\partial {\cal H}}}{{\partial r}},\\
\dot \psi  &= \frac{{\partial {\cal H}}}{{\partial {p_\psi }}},\quad {{\dot p}_\psi } =  - \frac{{\partial {\cal H}}}{{\partial \psi }}.
\end{aligned}
\end{equation}
For convenience of computation, we adopt the following units of length, mass and time \citep{jafari2023surfing}:
\begin{equation}\label{Eq10}
[L] = {a_s},\;[M] = \frac{{{m_p}{m_s}}}{{{m_p} + {m_s}}},\;[T] = \sqrt {{{{\cal G}\left( {{m_p} + {m_s}} \right)} \mathord{\left/
 {\vphantom {{{\cal G}\left( {{m_p} + {m_s}} \right)} {a_s^3}}} \right.
 \kern-\nulldelimiterspace} {a_s^3}}}.
\end{equation}
Under this system of normalized units, the reduced mass of system $m$, the gravitational parameter $\mu = {\cal G} (m_p + m_s)$ and the parameter ${\cal G}m_s m_p$ are all equal to unity in magnitude. From now on, we will discuss the spin-orbit coupling problem under the system of normalized units. For convenience, we denote the unit of length as $\rm LU$ and the unit of time as $\rm TU$.

In practice, we often take another set of variables to describe the translation and rotation, denoted by ${\bm Z} = (a,e,f,\phi-\varpi,\dot\phi)^{\rm T}$, where $a$ is the orbital semimajor axis, $e$ is the eccentricity, $f$ is the true anomaly and $\varpi$ is the longitude of pericenter. It is not difficult to realize mutual transformations among ${\bm X}$, ${\bm Y}$ and ${\bm Z}$ (to save space, the expressions of mutual transformation are not given here). 

In terms of ${\bm Z}$, the orbital angular momentum can be written as
\begin{equation}\label{Eq11}
p_\theta = m\sqrt{\mu a(1-e^2)} = \sqrt{a(1-e^2)}
\end{equation}
and the kinetic energy of system is given by
\begin{equation}\label{Eq12}
T = -\frac{{\cal G}m_p m_s}{2a} + \frac{1}{2}I_3{\dot\phi}^2 = -\frac{1}{2a} + \frac{1}{2}I_3{\dot\phi}^2.
\end{equation}
It is noted that normalized system of units are considered in Eqs. (\ref{Eq11}) and (\ref{Eq12}). 

Denote the ratio of second-to-primary mass as $\alpha_m = m_s/m_p <1$. The normalized mass of the primary and secondary can be specified by $\alpha_m$ in the following form,
\begin{equation}\label{Eq13}
1 < {m_s} = 1 + {\alpha _m} < 2,\quad 1 < {m_p} = 1 + \frac{1}{{{\alpha _m}}} < \infty,
\end{equation}
and the total angular momentum can be written as
\begin{equation}\label{Eq14}
{G_{\rm tot}} = \frac{1}{5}\left( {1 + {\alpha _m}} \right)\left( {1 + b_s^2} \right)\dot \phi  + \sqrt {a\left( {1 - {e^2}} \right)}.
\end{equation}
Considering the fact that the total angular momentum ${G_{\rm tot}}$ and the Hamiltonian $\cal H$ are conserved quantities of the current problem, we need to characterize them by taking advantage of equivalent parameters. 

For convenience, we take the reference semimajor axis $a_{\rm ref}$ to characterize the total angular momentum as follows:
\begin{equation}\label{Eq15}
{G_{\rm tot}} = \frac{1}{5}\left( {1 + {\alpha _m}} \right)\left( {1 + b_s^2} \right)\sqrt {{1 \mathord{\left/
 {\vphantom {1 {a_{ref}^3}}} \right.
 \kern-\nulldelimiterspace} {a_{\rm ref}^3}}}  + \sqrt {{a_{\rm ref}}}.
\end{equation}
It means that the parameter $a_{\rm ref}$ is the semimajor axis when the binary asteroid system is assumed at the nominal location of synchronous resonance (i.e., $\dot \phi = n$ with $n$ as the mean motion) under the assumption of zero eccentricity.

When the total angular momentum of system is provided by means of $a_{\rm ref}$, we further denote the eccentricity at the libration center of synchronous (1:1) spin-orbit resonance as the maximum eccentricity $e_{\max}$. Assuming the true anomaly at the nominal location of synchronous resonance as zero (pericentre), we can get the remaining variables at the resonance center: $r = \frac{a_{\rm ref}}{1+e_{\max}}$, $p_r = 0$, $\psi=0$ and $p_\psi =-\sqrt{a_{\rm ref}}$. As a result, the Hamiltonian can be characterized by $a_{\rm ref}$ and $e_{\max}$ as follows:
\begin{equation}\label{Eq16}
{\cal H}\left( {{G_{\rm tot}};r,{p_r},\psi ,{p_\psi }} \right) = {\left. {\cal H} \right|_{r = \frac{{{a_{\rm ref}}}}{{1 + {e_{\max }}}},{p_r} = 0,\psi  = 0,{p_\psi } =  - \sqrt{a_{\rm ref}} }}.
\end{equation}
According to Eqs. (\ref{Eq15}) and (\ref{Eq16}), the pair of parameters $(a_{\rm ref}, e_{\max})$ can be used to specify the total angular momentum and Hamiltonian.

\begin{figure*}
\centering
\includegraphics[width=0.66\columnwidth]{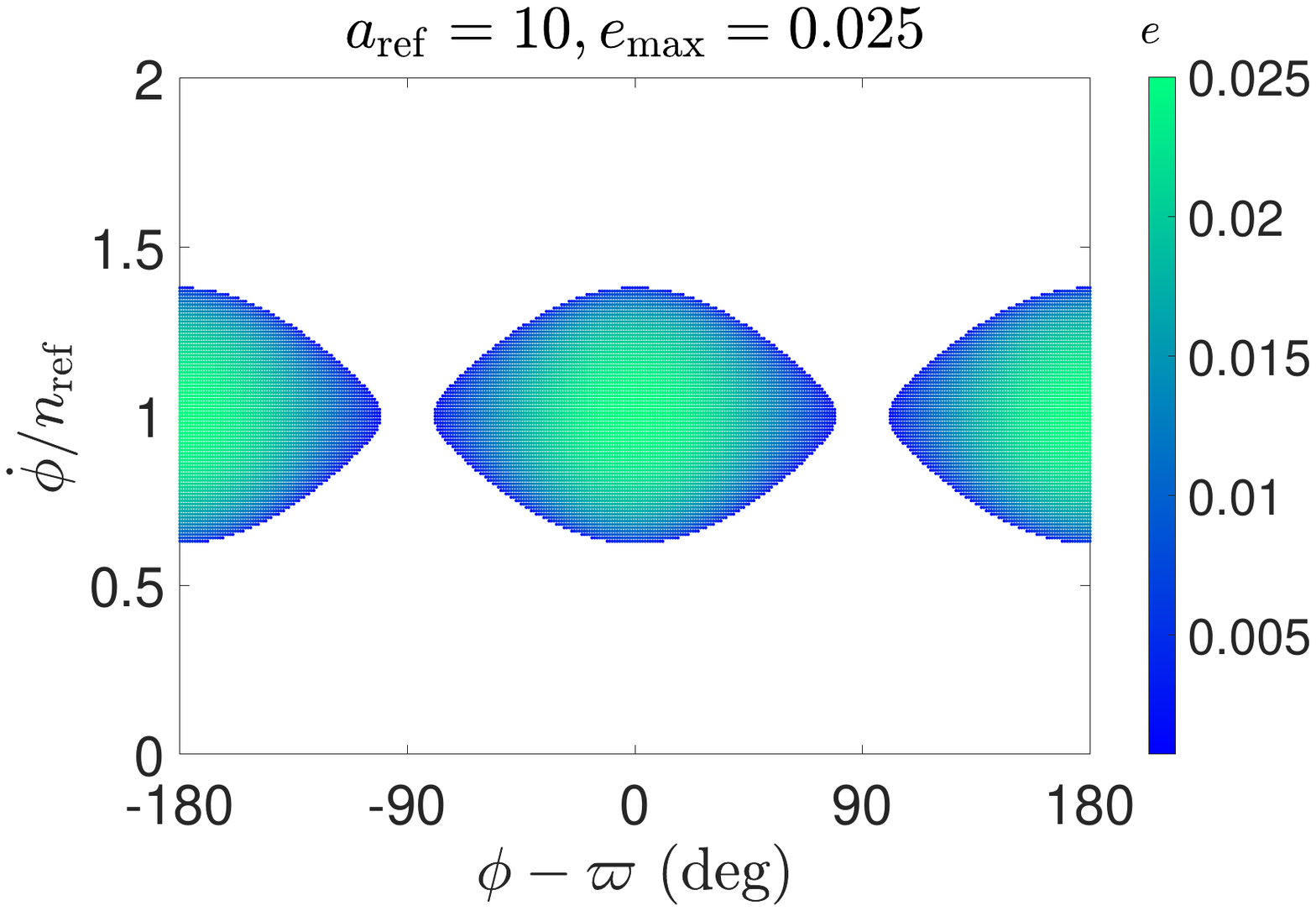}
\includegraphics[width=0.66\columnwidth]{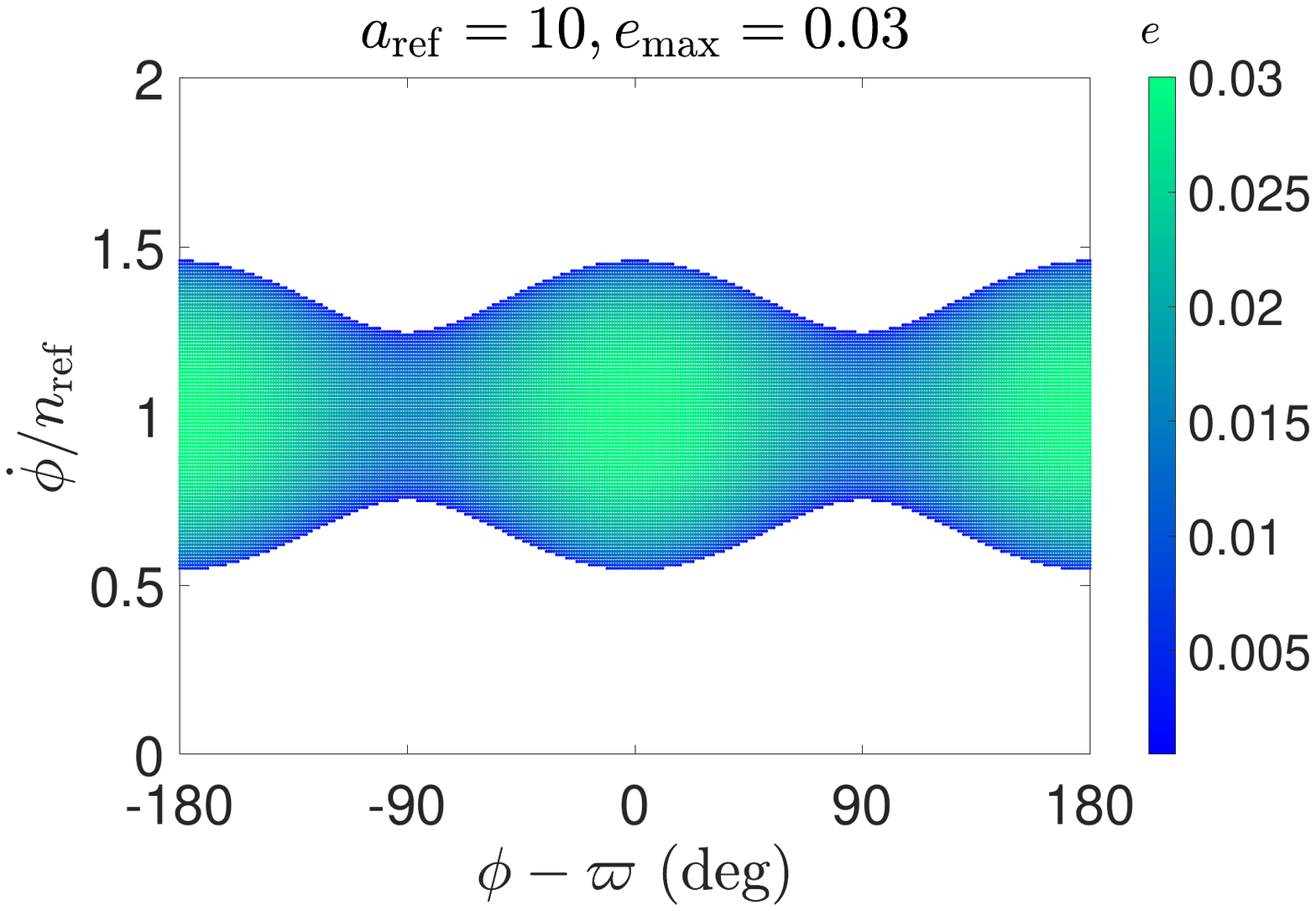}
\includegraphics[width=0.66\columnwidth]{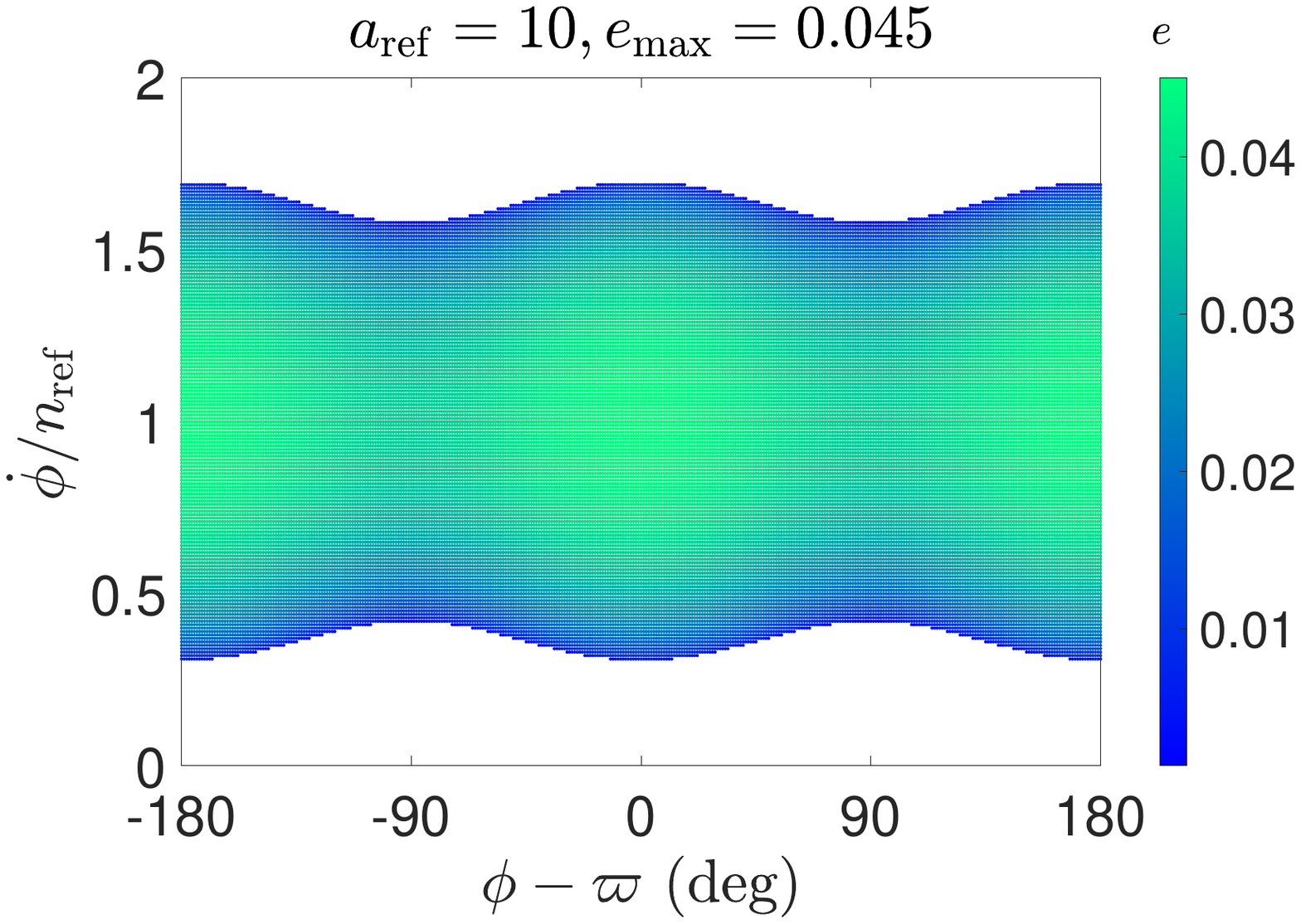}
\caption{Physically allowed regions corresponding to given Hamiltonian and total angular momentum, specified by the reference semimajor axis $a_{\rm ref}$ and the maximum eccentricity $e_{\max}$ (see Eqs. \ref{Eq15} and \ref{Eq16} for definition). The index shown in the color bar stands for the magnitude of eccentricity. For this example, the shape parameters of the secondary are taken as $a_s = 1$, $b_s = 0.95$ and $c_s = 0.85$ and the mass ratio is assumed at $\alpha_m = 0.05$.}
\label{Fig2}
\end{figure*}

The physically allowed regions (feasible regions) are controlled by the total angular momentum and Hamiltonian, which are specified by $(a_{\rm ref}, e_{\max})$. When $a_{\rm ref}$ is given, the total angular momentum is determined (see Eq. \ref{Eq15}). With a given $a_{\rm ref}$, the Hamiltonian is determined by the maximum eccentricity $e_{\max}$ (see Eq. \ref{Eq16}). Given the total angular momentum specified by $a_{\rm ref} = 10$, Fig. \ref{Fig2} shows the allowed regions at the Hamiltonian levels specified by $e_{\max}=0.025$, $e_{\max}=0.03$ and $e_{\max}=0.045$ (from the left to right). See the caption for detailed setting of system parameters. From Fig. \ref{Fig2}, we can observe that (a) the physically allowed area is greater when the maximum eccentricity $e_{\max}$ is larger, (b) the allowed region is symmetric with respect to the line of $\dot\phi = n_{\rm ref}$, (c) at the center of synchronous resonance the eccentricity reaches the maximum equal to $e_{\max}$, and (d) at the border the eccentricity approaches zero. Thus, when the total angular momentum specified by $a_{\rm ref}$ is given, the level of Hamiltonian specified by $e_{\max}$ determines the size of physically allowed region in the phase space.

\section{Poincar\'e sections}
\label{Sect3}

It is known that the spin-orbit coupling problem formulated in the previous section is a 2-DOF Hamiltonian model, depending on the total angular momentum. Regarding such a 2-DOF dynamical model, the technique of Poincar\'e sections is a powerful tool to explore global structures in the phase space. In general, continuous and smooth curves represent quasi-periodic (regular) solutions and those regions filled with scatter points are usually chaotic. In addition, the existence of islands in Poincar\'e sections are usually due to certain resonances and, in particular, the centre of island corresponds to a (stable) periodic orbit.

In this work, we define Poincar\'e sections by
\begin{equation}\label{Eq17}
{\rm mod} (f,2\pi) = 0
\end{equation}
which is also called the periapsis section, because the points on the defined section correspond to the moment at pericenter. In practice, we numerically integrate the equations of motion represented by Eq. (\ref{Eq9}) for a long enough period of time and record those points every time when the secondary passes through its pericenter (i.e., $f=0$). A similar definition of Poincar\'e section can be found in \citet{wisdom2004spin}, \citet{jafari2015widespread} and \citet{naidu2015near}.

As discussed before, the total angular momentum and Hamiltonian of system are determined by the reference semimajor axis $a_{\rm ref}$ and the maximum eccentricity $e_{\max}$. With given $(a_{\rm ref},e_{\max})$, Poincar\'e sections are shown in the $(\phi-\varpi,\dot\phi)$ space. According to the definition, the true anomaly for the points on sections is equal to zero. The remaining unknown variables including $a$ and $e$ can be solved from the given $G_{\rm tot}$ and $\cal H$. Thus, all the variables for the points on Poincar\'e sections are determinant.

Here we concentrate on the influences of four parameters upon numerical structures in Poincar\'e sections, including the shape parameter $b_s$, mass ratio $\alpha_m$, the reference semimajor axis $a_{\rm ref}$ and the maximum eccentricity $e_{\max}$. In each case we vary one parameter and fix the other parameters. The default parameters are taken as $(a_s,b_s,c_s)=(1,0.95,0.85)$, $\alpha_m = 0.05$, $a_{\rm ref} = 10$ and $e_{\max}=0.05$.

\begin{figure*}
\centering
\includegraphics[width=\columnwidth]{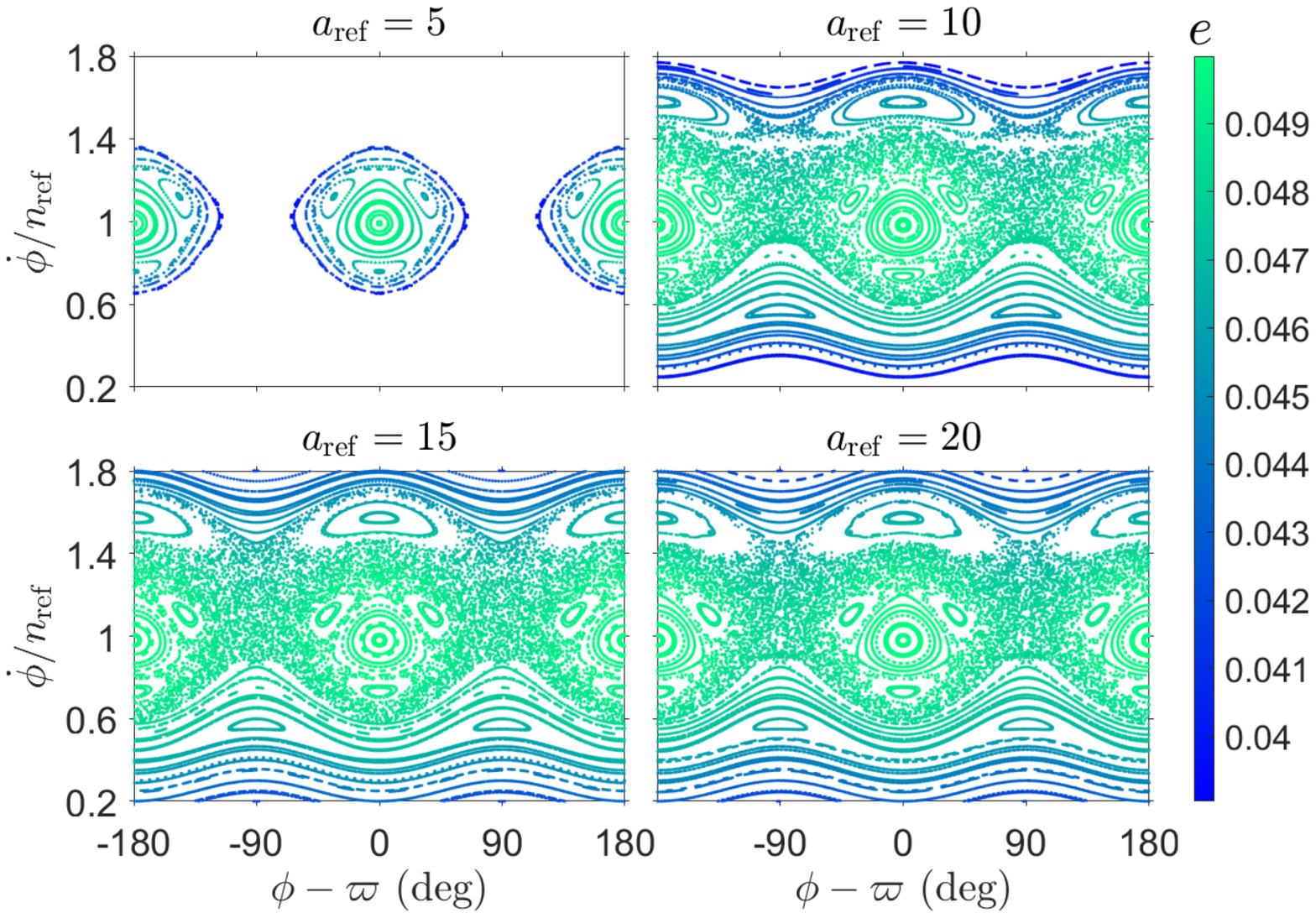}
\includegraphics[width=\columnwidth]{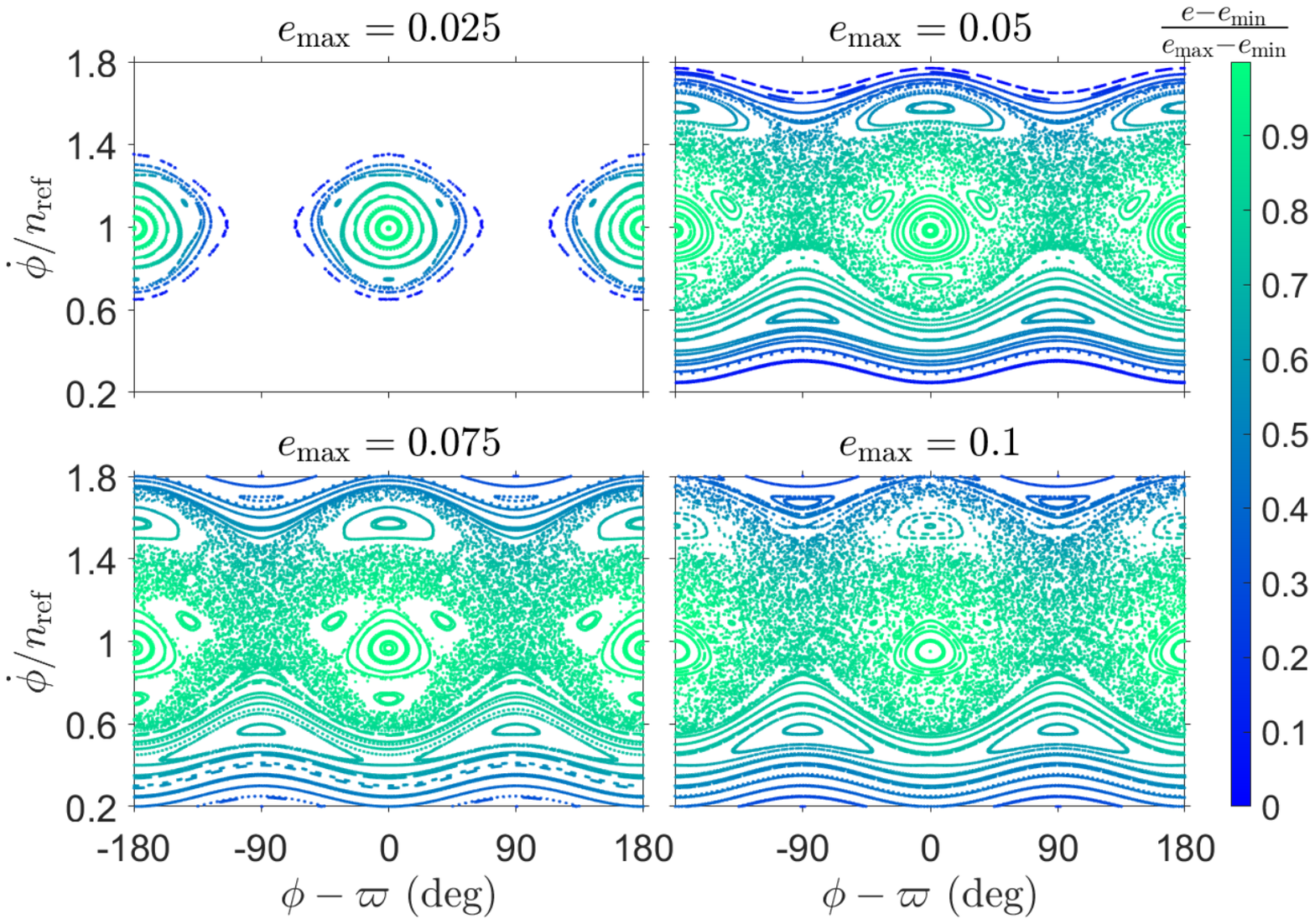}
\caption{Poincar\'e sections at different reference semimajor axes (\textit{left panel}) and at different maximum eccentricities (\textit{right panel}). The index shown in the colour bar stands for the magnitude of eccentricity.}
\label{Fig3}
\end{figure*}

In Fig. \ref{Fig3}, Poincar\'e sections for different reference semimajor axes at $a_{\rm ref} = 5$, $a_{\rm ref} = 10$, $a_{\rm ref} = 15$ and $a_{\rm ref} = 20$ are shown in the left panel with given $e_{\max}$ and the ones for different maximum eccentricities at $e_{\max} = 0.025$, $e_{\max} = 0.05$, $e_{\max} = 0.075$ and $e_{\max} = 0.1$ are presented in the right panel with given $a_{\rm ref}$. Because the reference semimajor axis $a_{\rm ref}$ determines the total angular momentum and ($a_{\rm ref}$, $e_{\max}$) determines the magnitude of Hamiltonian, we can state that results shown in Fig. \ref{Fig3} imply the influence of total angular momentum and Hamiltonian upon Poincar\'e sections.

As for the influence of $a_{\rm ref}$ (see the left panel of Fig. \ref{Fig3}), we can see that (a) the center of synchronous resonance is located at $2(\phi-\varpi) = 0$ and the saddle point is located at $2(\phi-\varpi) = \pi$ and (b) the allowed region is larger with a greater $a_{\rm ref}$. In the case of $a_{\rm ref} = 5$, the allowed region is small and it cannot cover the location of saddle point at $2(\phi-\varpi) = \pi$. The allowed region is filled with regular motion, corresponding to synchronous (1:1) spin-orbit resonance. Inside the island of synchronous (1:1) resonance, the secondary 3:1 spin-orbit resonance is visible. In the cases of $a_{\rm ref} = 10$, $a_{\rm ref} = 15$ and $a_{\rm ref} = 20$, the numerical structures arising in Poincar\'e sections are qualitatively similar. Firstly, different types of libration islands are visible and they correspond to the synchronous (1:1) resonance, secondary 3:1 resonance, and high-order resonances including the 2:1 and 2:3 spin-orbit resonances. Secondly, there are large chaotic regions, distributed around the islands of libration. 

As for the influence of $e_{\max}$ (see the right panel of Fig. \ref{Fig3}), we can see that the allowed region is larger with a higher $e_{\max}$. In the case of $e_{\max}=0.025$, the allowed region is small and it cannot cover the saddle point of the synchronous resonance. In this case, almost all the feasible region is filled with regular motion, corresponding to synchronous resonance. In addition, the secondary 3:1 spin-orbit resonance is visible. In the cases of $e_{\max} = 0.05$, $e_{\max} = 0.075$ and $e_{\max} = 0.1$, the Poincar\'e sections exhibit qualitatively similar structures. Firstly, islands of libration corresponding to the synchronous resonance, the secondary 3:1 resonance and high-order resonances can be observed. Secondly, chaotic seas can be observed around islands of libration and the area of chaotic region becomes larger when $e_{\max}$ is higher. Thirdly, the island of secondary 3:1 resonance becomes smaller and smaller as $e_{\max}$ increases.

\begin{figure*}
\centering
\includegraphics[width=\columnwidth]{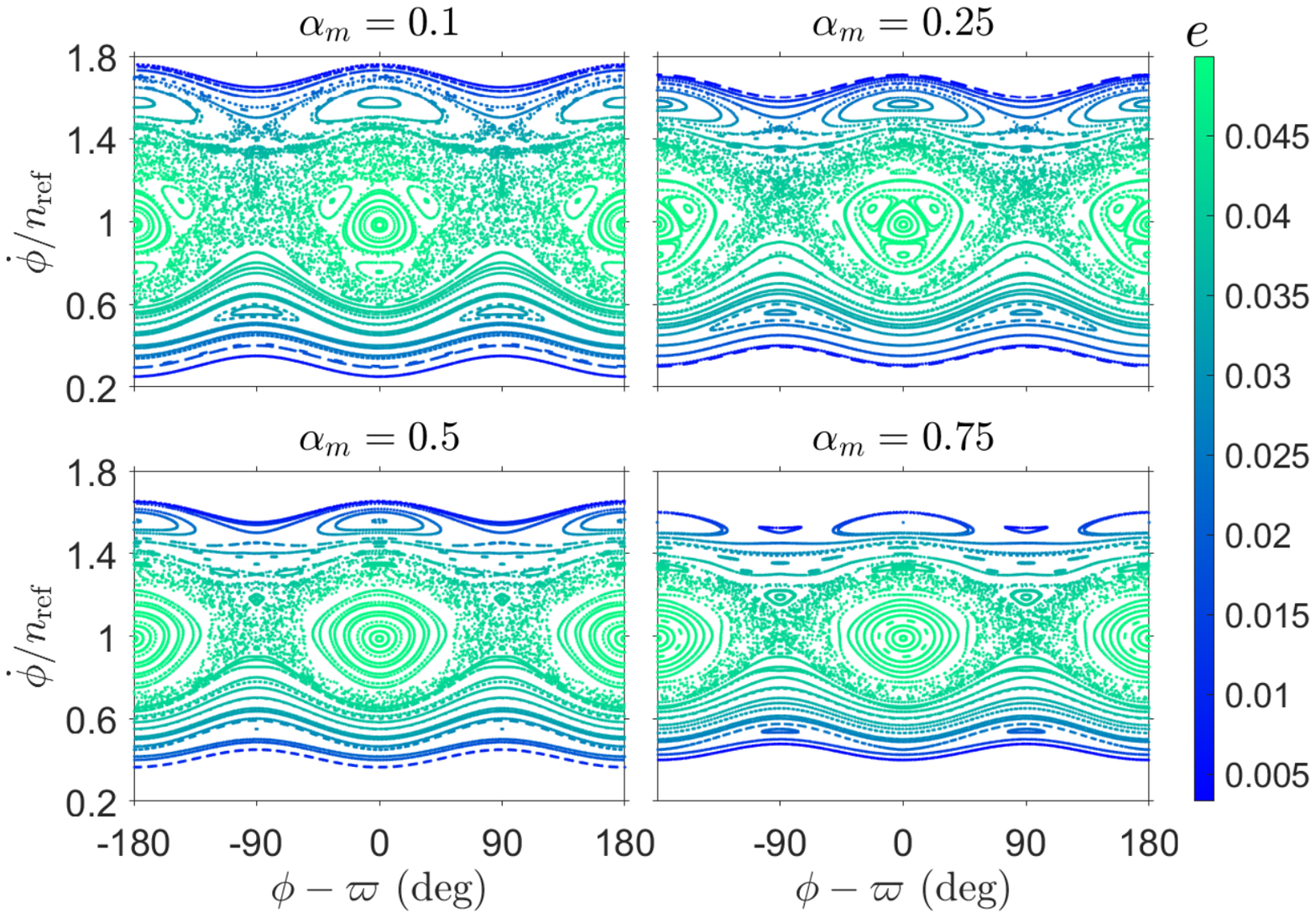}
\includegraphics[width=\columnwidth]{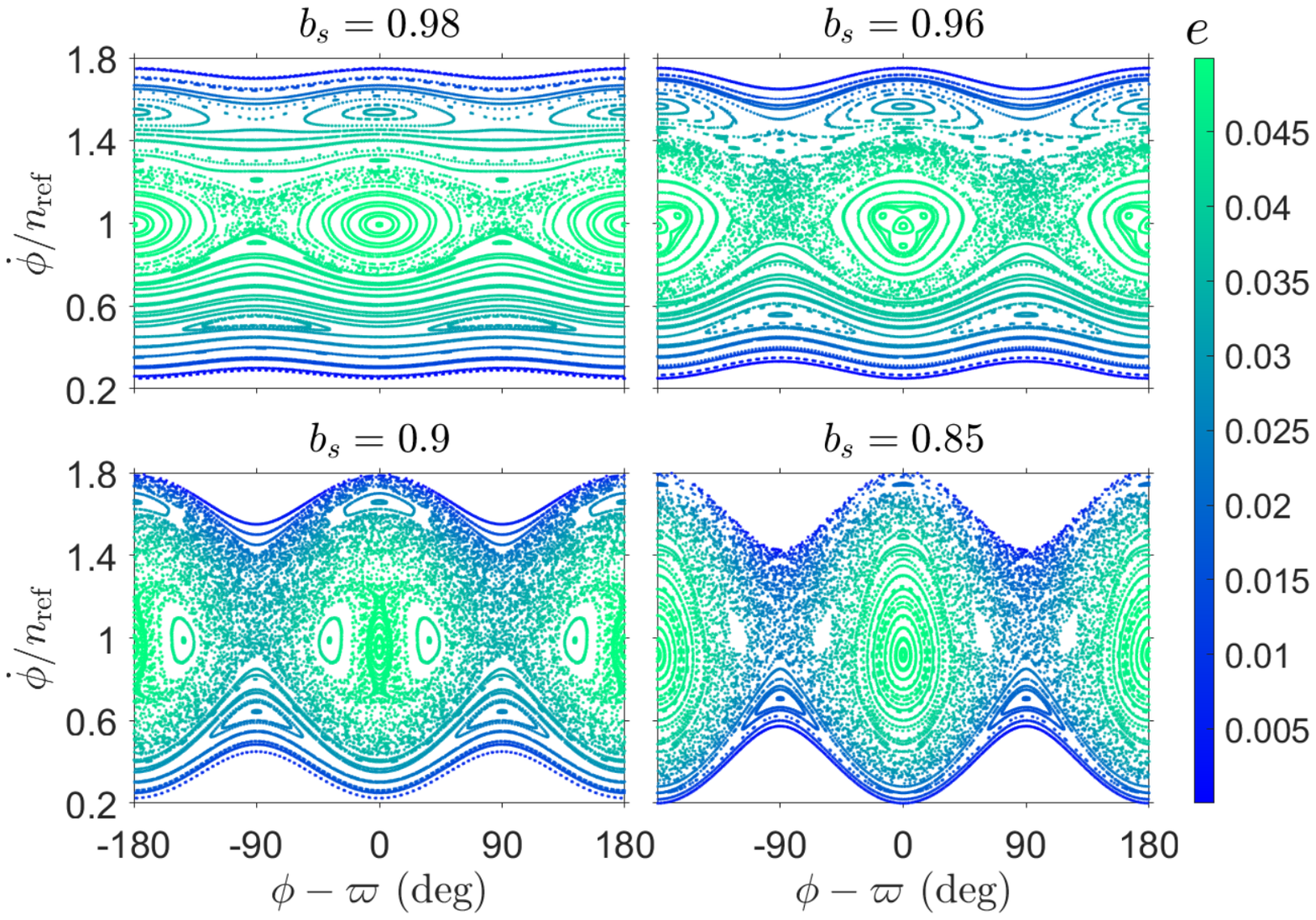}
\caption{Poincar\'e sections at different mass ratios (\textit{left panel}) and at different asphericity parameters (\textit{right panel}). The index shown in the color bar stands for the magnitude of eccentricity.}
\label{Fig4}
\end{figure*}

As for the influences of $\alpha_m$ (mass ratio) and $b_s$ (shape parameter) (see Fig. \ref{Fig4}), we can see that the feasible regions have no evident difference with different $\alpha_m$, while the feasible regions become smaller with the decreasing of $b_s$. A higher $\alpha_m$ indicates a larger mass of the secondary. A smaller $b_s$ means that the secondary has a larger equatorial elongation, indicating a higher degree of asphericity. In both cases, islands of synchronous (1:1) resonance, secondary resonances and high-order resonances can be observed. In addition, chaotic seas are observed around libration islands. However, dynamical structures associated with high-order and secondary resonances are significantly influenced by $\alpha_m$ and $b_s$. 

In the next section, analytical study is performed for the problem of spin-orbit coupling in order to understand dynamical structures arising in Poincar\'e sections.

\section{Perturbative treatments}
\label{Sect4}

In this section, we perform elliptic expansion for the Hamiltonian (\ref{Eq1}) and then study the spin-orbit coupling problem by taking advantage of perturbative treatments \citep{henrard1986perturbation,henrard1990semi}.

\subsection{Expansion of Hamiltonian function}
\label{Sect4-1}

In order to expand the Hamiltonian (\ref{Eq1}), the following elliptic expansion is used \citep{kaula1961analysis},
\begin{equation}\label{Eq18}
{\left( {\frac{a}{r}} \right)^l}\cos \left( {mf} \right) = \sum\limits_{s =  - \infty }^\infty  {X_s^{ - l,m}\left( e \right)\cos \left( {sM} \right)}
\end{equation}
where $M$ is the mean anomaly. Hansen coefficient $X_c^{a,b}(e)$ is a function of eccentricity and its lowest order in eccentricity is equal to $|b-c|$ \citep{murray1999solar}. After performing elliptic expansions, the Hamiltonian (\ref{Eq1}) takes the form,
\begin{equation}\label{Eq19}
\begin{aligned}
{\cal H} =&  - \frac{1}{{2a}} + \frac{{p_\phi ^2}}{{2{I_3}}}\\
& - \sum\limits_{s =  - \infty }^\infty  {\left\{ {\left[ { - \frac{{{C_{20}}}}{{2{a^3}}}X_s^{ - 3,0} + \frac{{3{C_{40}}}}{{8{a^5}}}X_s^{ - 5,0}} \right]\cos \left( {sM} \right)} \right.} \\
& + \left[ {\frac{{3{C_{22}}}}{{{a^3}}}X_s^{ - 3,2} - \frac{{15{C_{42}}}}{{2{a^5}}}X_s^{ - 5,2}} \right]\cos \left( {2\phi  - 2\varpi  - sM} \right)\\
&\left. { + \frac{{105{C_{44}}}}{{{a^5}}}X_s^{ - 5,4}\cos \left( {4\phi  - 4\varpi  - sM} \right)} \right\}.
\end{aligned}
\end{equation}
To formulate the Hamiltonian model, the following Delaunay's variables are introduced,
\begin{equation}\label{Eq20}
\begin{aligned}
l &= M,\quad L=\sqrt{a},\\
g &= \varpi,\quad G=\sqrt{a(1-e^2)}.
\end{aligned}
\end{equation}
Performing a canonical transformation,
\begin{equation}\label{Eq21}
\begin{aligned}
\psi_1 &= \phi-\varpi,\quad \Psi_1=p_\phi,\\
\psi_2 &= M,\quad \Psi_2 = \sqrt{a},\\
\psi_3 &= \varpi,\quad \Psi_3 = \sqrt{a(1-e^2)} + p_\phi = G_{\rm tot},
\end{aligned}
\end{equation}
we can rewrite Hamiltonian (\ref{Eq19}) in the following form:
\begin{equation}\label{Eq22}
\begin{aligned}
{\cal H} =&  - \frac{1}{{2\Psi _2^2}} + \frac{{\Psi _1^2}}{{2{I_3}}}\\
&  - \sum\limits_{n =  - \infty }^\infty  {\left\{ {\left[ { - \frac{{{C_{20}}}}{{2\Psi _2^6}}X_n^{ - 3,0} + \frac{{3{C_{40}}}}{{8\Psi _2^{10}}}X_n^{ - 5,0}} \right]\cos \left( {n{\psi _2}} \right)} \right.} \\
&+ \left[ {\frac{{3{C_{22}}}}{{\Psi _2^6}}X_n^{ - 3,2} - \frac{{15{C_{42}}}}{{2\Psi _2^{10}}}X_n^{ - 5,2}} \right]\cos \left( {2{\psi _1} - n{\psi _2}} \right)\\
&+\left. {\frac{{105{C_{44}}}}{{\Psi _2^{10}}}X_n^{ - 5,4}\cos \left( {4{\psi _1} - n{\psi _2}} \right)} \right\}
\end{aligned}
\end{equation}
where the Hansen coefficients $X_c^{a,b}$ are related to the eccentricity $e$, which can be expressed by canonical variables. As the angular coordinate $\psi_3$ is absent from the Hamiltonian (\ref{Eq7}), its conjugate moment $\Psi_3 = G_{\rm tot}$ is a motion integral. Under such a 2-DOF Hamiltonian model, the pair of variables $(\psi_1,\Psi_1)$ stands for the degree of freedom associated with rotational motion and the pair $(\psi_2,\Psi_2)$ represent the degree of freedom associated with orbital motion. In practice, we need to truncate the Hamiltonian expansion (\ref{Eq22}) up to a certain order in eccentricity $e$ according to the requirement of accuracy.

It is known that in the spin-orbit coupling problem the synchronous (1:1) resonance is the strongest one among varieties of spin-orbit resonances. To isolate the argument associated with synchronous spin-orbit resonance, we further introduce the following canonical transformation,
\begin{equation}\label{Eq23}
\begin{aligned}
{\sigma _1} &= {\psi _1} - {\psi _2},\quad {\Sigma _1} = {\Psi _1},\\
{\sigma _2} &= {\psi _2},\quad {\Sigma _2} = {\Psi _2} + {\Psi _1},\\
{\sigma _3} &= {\psi _3},\quad {\Sigma _3} = {\Psi _3} \equiv {G_{\rm tot}}.
\end{aligned}
\end{equation}
Then, the Hamiltonian (\ref{Eq22}) can be re-organized as follows:
\begin{equation}\label{Eq24}
\begin{aligned}
{\cal H} =&  - \frac{1}{{2{{\left( {{\Sigma _2} - {\Sigma _1}} \right)}^2}}} + \frac{{\Sigma _1^2}}{{2{I_3}}}\\
&  - \sum\limits_{n =  - \infty }^\infty  {\left\{ {\left[ { - \frac{{{C_{20}}X_n^{ - 3,0}}}{{2{{\left( {{\Sigma _2} - {\Sigma _1}} \right)}^6}}} + \frac{{3{C_{40}}X_n^{ - 5,0}}}{{8{{\left( {{\Sigma _2} - {\Sigma _1}} \right)}^{10}}}}} \right]\cos \left( {n{\sigma _2}} \right)} \right.} \\
& + \left[ {\frac{{3{C_{22}}X_n^{ - 3,2}}}{{{{\left( {{\Sigma _2} - {\Sigma _1}} \right)}^6}}} - \frac{{15{C_{42}}X_n^{ - 5,2}}}{{2{{\left( {{\Sigma _2} - {\Sigma _1}} \right)}^{10}}}}} \right]\cos \left[ {2{\sigma _1} + \left( {2 - n} \right){\sigma _2}} \right]\\
&\left. { + \frac{{105{C_{44}}X_n^{ - 5,4}}}{{{{\left( {{\Sigma _2} - {\Sigma _1}} \right)}^{10}}}}\cos \left[ {4{\sigma _1} + \left( {4 - n} \right){\sigma _2}} \right]} \right\}.
\end{aligned}
\end{equation}
We can see that $\sigma_1 = \phi - \varpi - M$ is the argument for the synchronous (1:1) spin-orbit resonance, and $\sigma_2 = M$ stands for the translational periodic motion. In particular, $\sigma_3 = \varpi$ is a cyclic variable in the Hamiltonian, thus its conjugate momentum $\Sigma_3 \equiv {G_{\rm tot}}$ is a motion integral.

From the viewpoint of perturbative treatment \citep{henrard1990semi}, the Hamiltonian (\ref{Eq24}) is composed of a kernel (unperturbed) Hamiltonian denoted by ${\cal H}_0$ and a perturbation part denoted by ${\cal H}_1$, given by
\begin{equation}\label{Eq25}
{\cal H} = {\cal H}_0 + \varepsilon {\cal H}_1 
\end{equation}
where the expressions of ${\cal H}_0$ and $\varepsilon{\cal H}_1$ are to be discussed in the coming section.

\subsection{The synchronous spin-orbit resonance}
\label{Sect4-2}

In Hamiltonian (\ref{Eq24}), we keep Hansen coefficients with the lowest order of $e^0$, leading to the kernel Hamiltonian as follows:
\begin{equation}\label{Eq26}
\begin{aligned}
{{\cal H}_0^{(1)}} =&  - \frac{1}{{2{{\left( {{\Sigma _2} - {\Sigma _1}} \right)}^2}}} + \frac{{\Sigma _1^2}}{{2{I_3}}} + \frac{{{C_{20}}X_0^{ - 3,0}}}{{2{{\left( {{\Sigma _2} - {\Sigma _1}} \right)}^6}}} - \frac{{3{C_{40}}X_0^{ - 5,0}}}{{8{{\left( {{\Sigma _2} - {\Sigma _1}} \right)}^{10}}}}\\
& - \left[ {\frac{{3{C_{22}}X_2^{ - 3,2}}}{{{{\left( {{\Sigma _2} - {\Sigma _1}} \right)}^6}}} - \frac{{15{C_{42}}X_2^{ - 5,2}}}{{2{{\left( {{\Sigma _2} - {\Sigma _1}} \right)}^{10}}}}} \right]\cos 2{\sigma _1}\\
&  - \frac{{105{C_{44}}}}{{{{\left( {{\Sigma _2} - {\Sigma _1}} \right)}^{10}}}}X_4^{ - 5,4}\cos 4{\sigma _1}.
\end{aligned}
\end{equation}
Usually, the eccentricity of mutual orbit is very small ($e < 0.05$). For simplicity, removing those terms with order in eccentricity greater than zero from ${\cal H}_0^{(1)}$, we could obtain the second kernel Hamiltonian, given by
\begin{equation}\label{Eq27}
\begin{aligned}
{\cal H}_0^{\left( 2 \right)} =&  - \frac{1}{{2{{\left( {{\Sigma _2} - {\Sigma _1}} \right)}^2}}} + \frac{{\Sigma _1^2}}{{2{I_3}}} + \frac{{{C_{20}}}}{{2{{\left( {{\Sigma _2} - {\Sigma _1}} \right)}^6}}} - \frac{{3{C_{40}}}}{{8{{\left( {{\Sigma _2} - {\Sigma _1}} \right)}^{10}}}}\\
&- \left[ {\frac{{3{C_{22}}}}{{{{\left( {{\Sigma _2} - {\Sigma _1}} \right)}^6}}} - \frac{{15{C_{42}}}}{{2{{\left( {{\Sigma _2}
- {\Sigma _1}} \right)}^{10}}}}} \right]\cos 2{\sigma _1}\\
&- \frac{{105{C_{44}}}}{{{{\left( {{\Sigma _2} - {\Sigma _1}} \right)}^{10}}}}\cos 4{\sigma _1}.
\end{aligned}
\end{equation}
Usually, the 4th order and degree harmonics coefficients $C_{40}$ and $C_{44}$ are much smaller than the second order and degree coefficients in magnitude. Thus, we could further remove those terms associated with $C_{40}$ and $C_{44}$ to get the third kernel Hamiltonian as follows:
\begin{equation}\label{Eq28}
\begin{aligned}
{\cal H}_0^{\left( 3 \right)} =&  - \frac{1}{{2{{\left( {{\Sigma _2} - {\Sigma _1}} \right)}^2}}} + \frac{{\Sigma _1^2}}{{2{I_3}}} + \frac{{{C_{20}}}}{{2{{\left( {{\Sigma _2} - {\Sigma _1}} \right)}^6}}}\\
& - \frac{{3{C_{22}}}}{{{{\left( {{\Sigma _2} - {\Sigma _1}} \right)}^6}}}\cos 2{\sigma _1}.
\end{aligned}
\end{equation}
Furthermore, the secular term associated with $C_{20}$ is usually much smaller than the other two terms. Thus, we could further remove the term associated with $C_{20}$ to reach the fourth kernel Hamiltonian,
\begin{equation}\label{Eq29}
{\cal H}_0^{\left( 4 \right)} =  - \frac{1}{{2{{\left( {{\Sigma _2} - {\Sigma _1}} \right)}^2}}} + \frac{{\Sigma _1^2}}{{2{I_3}}} - \frac{{3{C_{22}}}}{{{{\left( {{\Sigma _2} - {\Sigma _1}} \right)}^6}}}\cos 2{\sigma _1}.
\end{equation}
Under different kernel Hamiltonian models, the perturbation part is denoted by
\begin{equation}\label{Eq30}
{\cal H}_1^{(i)} = {\cal H} - {\cal H}_0^{(i)}, \quad i=1,2,3,4.
\end{equation}

\begin{figure*}
\centering
\includegraphics[width=\columnwidth]{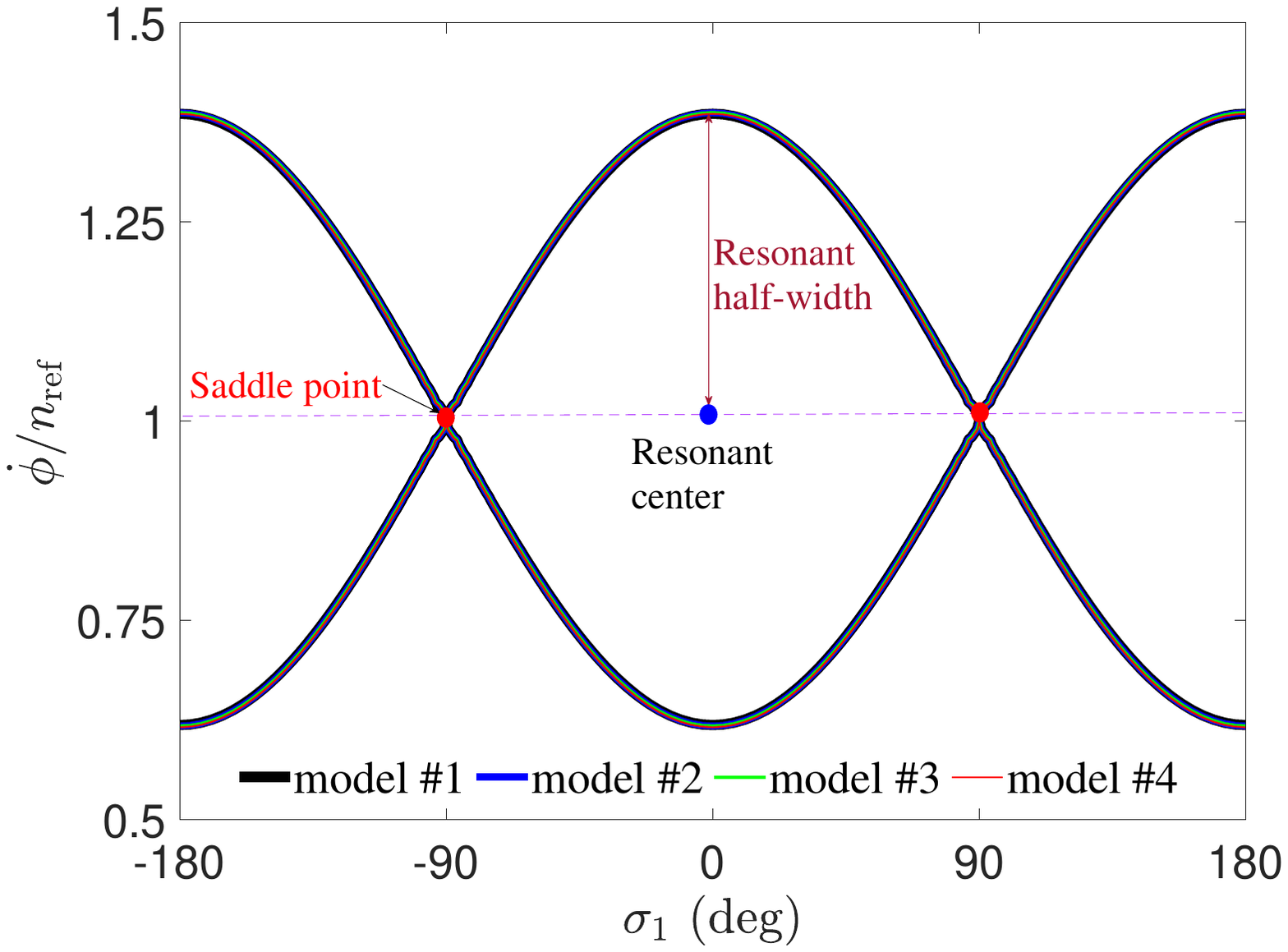}
\includegraphics[width=\columnwidth]{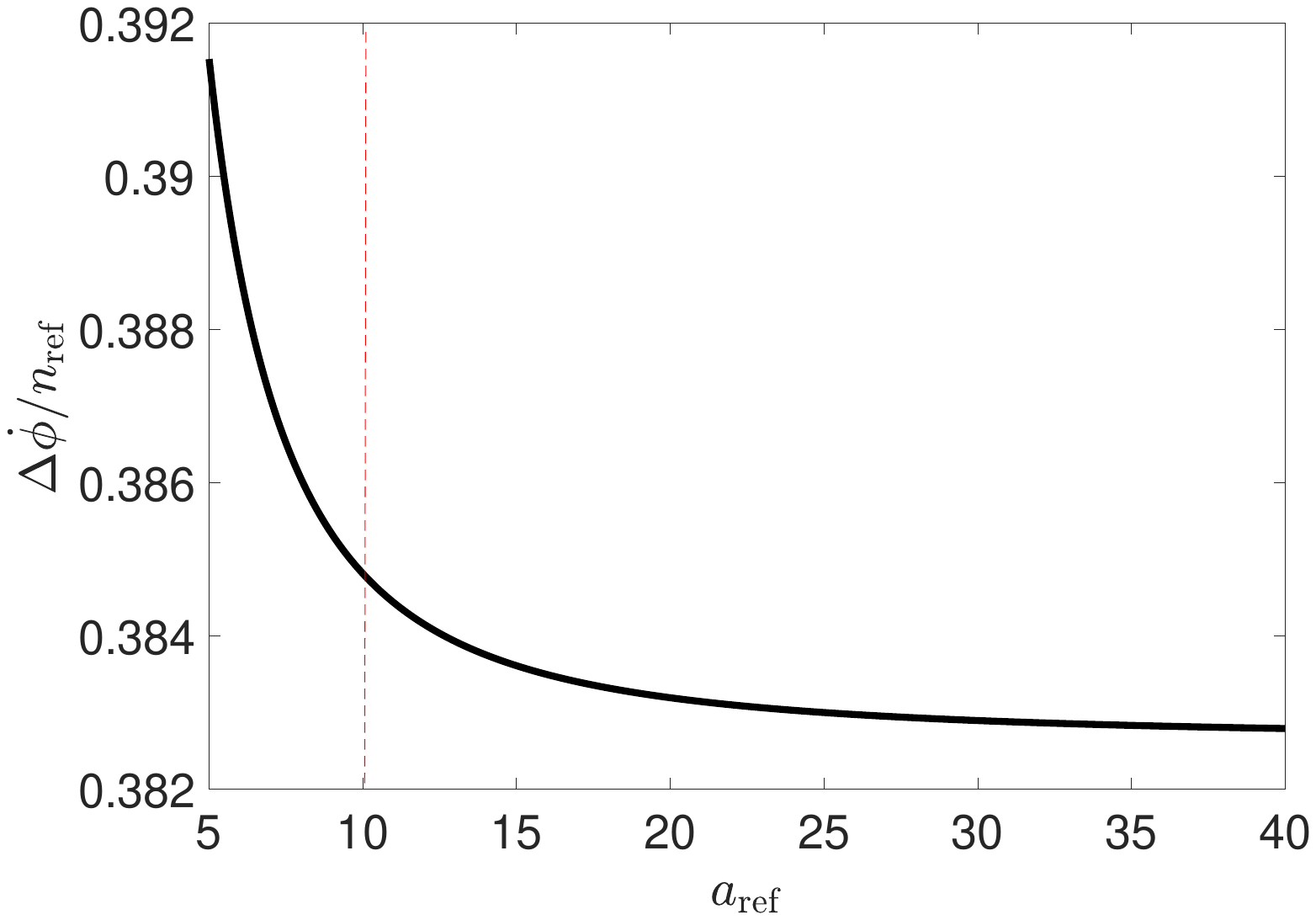}
\caption{Comparison of dynamical separatrices under different kernel Hamiltonian models (\textit{left panel}) and resonant half-width of synchronous spin-orbit resonance as a function of $a_{\rm ref}$ determined under Hamiltonian model \#4 (\textit{right panel}). The location of resonant center and saddle point is marked. Red dashed line shown in the right panel shows the location of $a_{\rm ref} = 10$. Dynamical separatrix divides the entire phase space into regions of libration and circulation in terms of synchronous (1:1) spin-orbit resonance. The default parameters of system are set as $a_s = 1$, $b_s = 0.95$, $c_s = 0.85$, $\alpha_m = 0.05$, $a_{\rm ref}=10$ and $e_{\max}=0.05$.}
\label{Fig5}
\end{figure*}

Dynamical separatrices under four kernel Hamiltonian models (denoted by models \#1, \#2, \#3, \#4) are presented in the left panel of Fig. \ref{Fig5}. Please refer to the caption for detailed setting of parameters. It is observed that these four kernel Hamiltonian models are in good agreement for describing phase-space structures of synchronous spin-orbit resonance. In particular, the libration center (marked by solid blue circle) is located at $(2\sigma_1=0,\dot\phi/n_{\rm ref}=1.0)$ and the saddle point (marked by solid red circle) is located at $(2\sigma_1=\pi,\dot\phi/n_{\rm ref}=1.0)$.

For the purpose of simplicity, we take the fourth kernel Hamiltonian in the following discussion and denote it as ${\cal H}_0 = {\cal H}_0^{(4)}$. Here, we derive the resonant half-width of the synchronous spin-orbit resonance based on pendulum approximation \citep{murray1999solar}. Expanding the secular Hamiltonian around the center $(2\sigma_1 = 0, \Sigma_{1,\rm res})$ up to the second order in $\Delta \Sigma_1$, we can arrive at the pendulum approximation for the unperturbed Hamiltonian, 
\begin{equation}\label{Eq31}
\begin{aligned}
\Delta {{\cal H}_{1:1}} =& \frac{1}{2}\left( {\frac{1}{{{I_3}}} - \frac{3}{{\left( {{\Sigma _2} - {\Sigma _1}} \right)_{\rm res}^4}}} \right)\Delta \Sigma _1^2\\
&- \frac{{3{C_{22}}}}{{\left( {{\Sigma _2} - {\Sigma _1}} \right)_{\rm res}^6}}\cos 2{\sigma _1}.
\end{aligned}
\end{equation}
Replacing $\left( {{\Sigma _2} - {\Sigma _1}} \right)_{\rm res}^2 = a_{\rm ref}$ in Eq. (\ref{Eq31}), we can simplify the Hamiltonian as follows:
\begin{equation}\label{Eq32}
\Delta {{\cal H}_{1:1}} = \frac{1}{2}\left( {\frac{1}{{{I_3}}} - \frac{3}{{{a_{\rm ref}^2}}}} \right)\Delta \Sigma _1^2 - \frac{{3{C_{22}}}}{{{a_{\rm ref}^3}}}\cos 2{\sigma _1}.
\end{equation}
According to the definition of dynamical separatrix (level curves of Hamiltonian passing through saddle point), we can get its explicit expression in the $(\sigma_1,\Delta \Sigma_1)$ space, 
\begin{equation}\label{Eq32-1}
\frac{{6{C_{22}}}}{{{a_{\rm ref}^3}}} \left( 1+  \cos 2{\sigma _1}\right) =\left( {\frac{1}{{{I_3}}} - \frac{3}{{{a_{\rm ref}^2}}}} \right)\Delta \Sigma _1^2.
\end{equation}
From the expression of separatrix, we can see that $|\Delta \Sigma_1|$ takes its minimum at the saddle point ($2\sigma_1=\pi$) and takes its maximum at the resonance center ($2\sigma_1=0$). The maximum value of $|\Delta \Sigma_1|$ is usually referred to as the resonant half-width (see Fig. \ref{Fig5}), which can be used to measure the size of libration region. 

Thus, the resonant half-width in terms of $\Delta {\Sigma _1}$ and $\Delta \dot \phi$, measuring the strength of synchronous spin-orbit resonance, can be expressed as
\begin{equation}\label{Eq33}
\Delta {\Sigma _1} = \sqrt {\frac{{12{C_{22}}}}{{\frac{{{a_{\rm ref}^3}}}{{{I_3}}} - 3a_{\rm ref}}}}  \Rightarrow {\left( {\Delta \dot \phi /n_{\rm ref}} \right)_{1:1}}{ = }\sqrt {\frac{{12{C_{22}}}}{{{I_3}\left( {1 - \frac{{3{I_3}}}{{{a_{\rm ref}^2}}}} \right)}}}.
\end{equation}
It shows that the resonant half-width is dependent on the reference semimajor axis $a_{\rm ref}$ (the total angular momentum), $I_3$ and $C_{22}$. The latter two parameters are determined by the mass ratio $\alpha_m$ and the shape parameter $b_s$. As a result, the resonant half-width is a function of $a_{\rm ref}$, $\alpha_m$ and $b_s$. The right panel of Fig. \ref{Fig5} shows the variation of resonant half-width as a function of $a_{\rm ref}$. Please see the caption for the detailed setting of system parameters. It is observed that the resonant half-width decreases with the reference semimajor axis $a_{\rm ref}$. 

In particular, when the reference semimajor axis approaches infinity ($a_{\rm ref} \to \infty$), the orbital angular momentum is much larger than the rotational angular momentum, in this case the influence of rotation upon the translational motion can be ignored, i.e., the orbit of the secondary around the primary remains invariant \citep{murray1999solar}. As a result, the current spin-orbit coupling can be reduced to the problem of spin-orbit resonance (a restricted configuration). Under the restricted configuration, it holds $\frac{3I_3}{a^2} \to 0$, leading to a simplified expression of resonant half-width as follows:
\begin{equation}\label{Eq34}
{\left( {\Delta \dot \phi /n} \right)_{a_{\rm ref} \to \infty}} = \sqrt {\frac{{12{C_{22}}}}{{{I_3}}}}
\end{equation}
which can be further written as
\begin{equation}\label{Eq35}
{\left( {\Delta \dot \phi /n} \right)_{a_{\rm ref} \to \infty}} = \sqrt {\frac{{3\left( {1 - b_s^2} \right)}}{{\left( {1 + {\alpha _m}} \right)\left( {1 + b_s^2} \right)}}}.
\end{equation}
It is observed that ${\left( {\Delta \dot \phi /n} \right)_{a_{\rm ref} \to \infty}}$ is a decreasing function of $b_s$ and, in particular, it is equal to zero in the limit case of $b_s = 1$, corresponding to zero equatorial elongation. Furthermore, under the planet--satellite system, the mass ratio holds $\alpha_m \to 0$ (a limit configuration), thus the resonant half-width further becomes
\begin{equation}\label{Eq36}
{\left( {\Delta \dot \phi /n} \right)_{a_{\rm ref} \to \infty, \alpha_m \to 0}} = \sqrt {\frac{{3\left( {1 - b_s^2} \right)}}{{{1 + b_s^2}}}},
\end{equation}
which is equal to the asphericity parameter $\alpha$ of the secondary \citep{wisdom1984chaotic,murray1999solar}. It means that, under the restricted case, the asphericity parameter $\alpha$ stands for the resonant half-width (or strength) of the synchronous spin-orbit resonance \citep{lei2024dynamical}. In other words, the synchronous spin-orbit resonance is stronger if the secondary object has a higher equatorial elongation.

\subsection{Introduction of action-angle variables}
\label{Sect4-3}

\begin{figure*}
\centering
\includegraphics[width=\columnwidth]{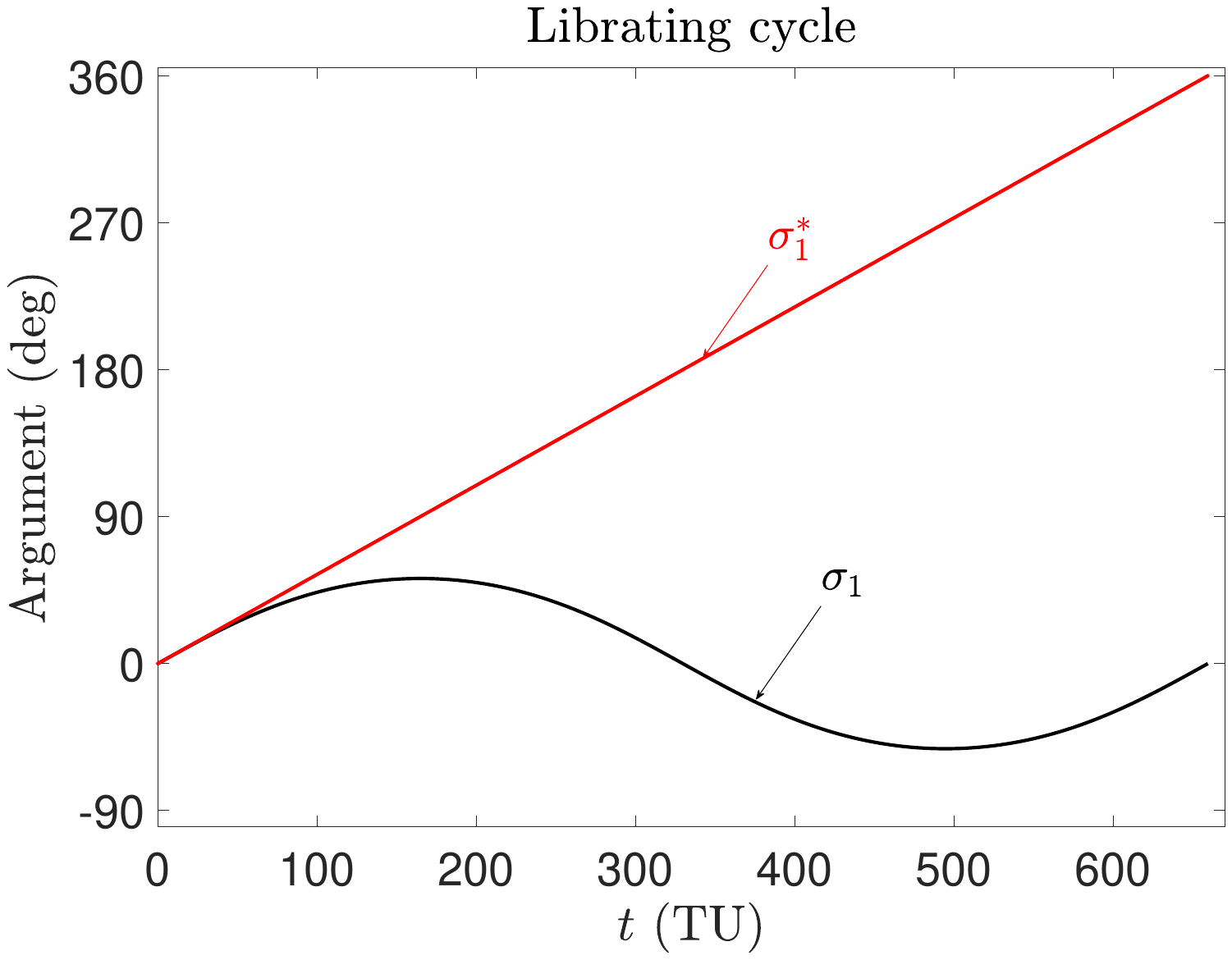}
\includegraphics[width=\columnwidth]{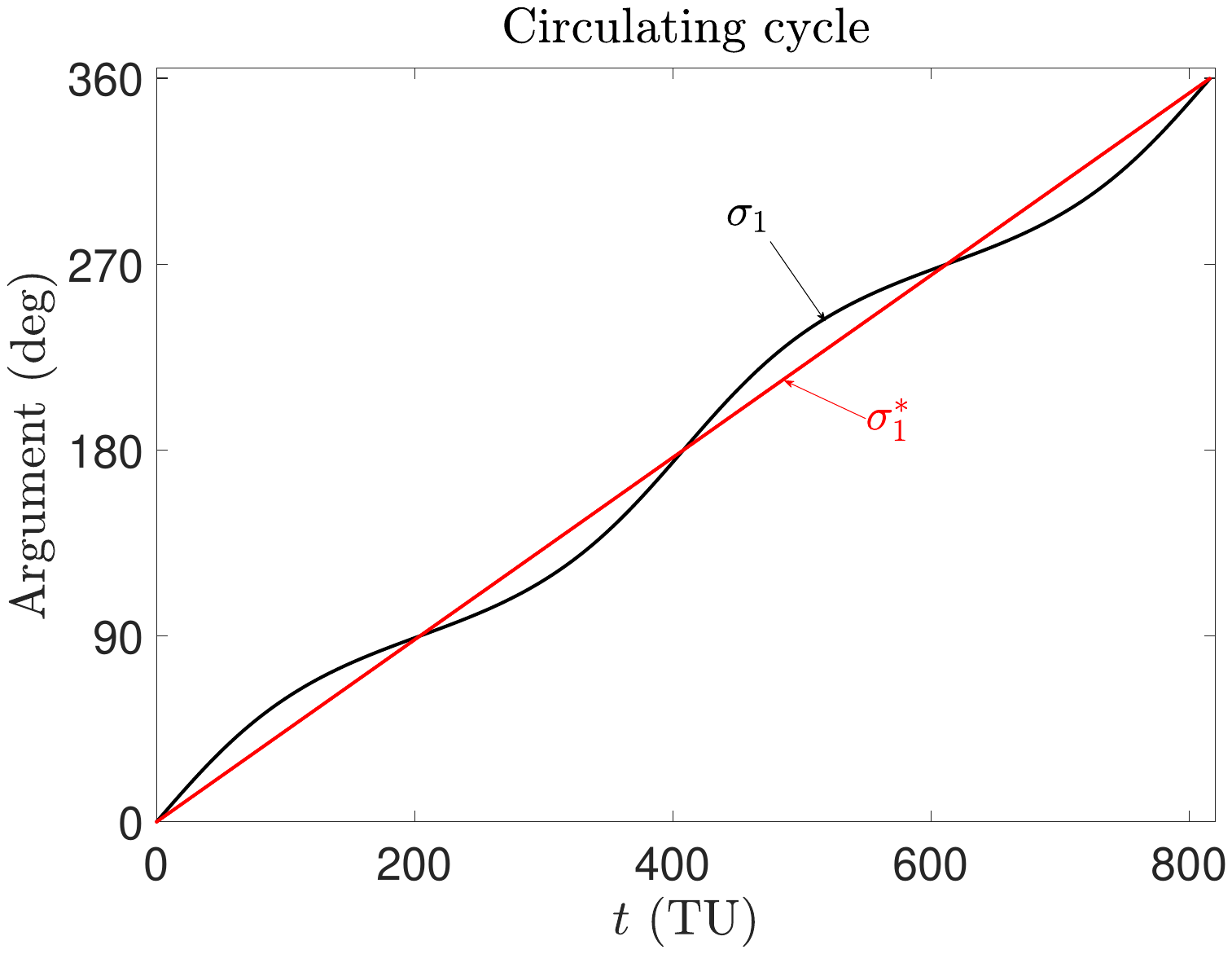}
\caption{Time evolution of the old and new angular coordinates (i.e., before and after canonical transformation) for a librating cycle (\textit{left panel}) and a circulating cycle (\textit{right panel}).}
\label{Fig6}
\end{figure*}

The unperturbed Hamiltonian given by Eq. (\ref{Eq29}) determines an integrable model, where $\Sigma_2$ is a motion integral. Under the integrable Hamiltonian, it is possible to introduce a set of action-angle variables as follows \citep{morbidelli2002modern}:
\begin{equation}\label{Eq37}
\begin{aligned}
\sigma _1^{*} &= {\sigma _1} - {\rho _1} = \frac{{2\pi }}{{{T_1}}}t,\quad \Sigma _1^{*} = \frac{1}{{2\pi }} \oint {{\Sigma _1}{\rm d}{\sigma _1}},\\
\sigma _2^{*} &= {\sigma _2} - {\rho _2} = {\sigma _2^* (0)} + \frac{{2\pi }}{{{T_2}}}t,\quad \Sigma _2^{*} = {\Sigma _2}
\end{aligned}
\end{equation}
where $T_1$ and $T_2$ are periods of $\sigma_1$ and $\sigma_2$ for a librating and/or circulating cycle under the unperturbed Hamiltonian model, and ${\rho _1}$ and ${\rho _2}$ are periodic functions with zero average and they hold the same period ($T_1$) of $\sigma_1$. At the initial instant, it holds $\sigma_1^*(0) = 0$ and $\sigma_2^*(0) = \sigma_2(0)$. After transformation, it is observed that the new angular coordinates are linear functions of time, as shown in Fig. \ref{Fig6}.

The transformation given by Eq. (\ref{Eq37}) is canonical with the following generating function,
\begin{equation}\label{Eq38}
{\cal S}\left( {{\sigma _1},{\sigma _2},\Sigma _1^*,\Sigma _2^*} \right) = {\sigma _2}\Sigma _2^* + \int {{\Sigma _1}{\rm d}{\sigma _1}}
\end{equation}

In the following simulations, the direct and inverse transformations between the set of variables $\left(\sigma_1,\sigma_2,\Sigma_1,\Sigma_2\right)$ and the set of transformed variables $\left(\sigma_1^*,\sigma_2^*,\Sigma_1^*,\Sigma_2^*\right)$ are often used. Thus, here let us provide some discussions about them under the unperturbed Hamiltonian (\ref{Eq29}). For both the librating and circulating cycles, it is assumed to be initially located at the point with $2\sigma_1 = 0^{\circ}$ (corresponding to the center of synchronous spin-orbit resonance). In particular, for the librating cycle it is further assumed to be initially located at the point with $\dot\phi>n_{\rm ref}$.

According to Eq. (\ref{Eq37}), we can get $\Sigma_2^* = \Sigma_2$. Thus, for direct transformation, we only need to determine $\sigma_{1,2}^*$ and $\Sigma_1^*$ through the following steps.
\begin{itemize}
\item Step I. Performing inverse integration under the unperturbed Hamiltonian model to reach the starting point, we can determine the initial state $\left(\sigma_1(0) = 0,\sigma_2(0),\Sigma_1(0),\Sigma_2(0)\right)$ as well as the time $t$. According to the discussion made about Eq. (\ref{Eq37}), we could get $\sigma_1^* (0) =0$ and $\sigma_2^* (0) = \sigma_2(0)$.
\item Step II. Integrating the equation of motion about the unperturbed Hamiltonian over one period of $\sigma_1$, it is possible to determine the periods ($T_1$ for $\sigma_1$ and $T_2$ for $\sigma_2$) as well as $\Sigma_1^*$.
\item Step III. According to Eq. (\ref{Eq37}), it becomes possible to determine $\sigma_1^*$ and $\sigma_2^*$. 
\end{itemize}
For the inverse transformation, it is known that $\Sigma_2 = \Sigma_2^*$, thus we only need to determine $\sigma_{1,2}$ and $\Sigma_1$. The detailed procedures are given as follows:
\begin{itemize}
\item Step I. At the initial time $t=0$, it is known that $\sigma_1(0) = 0$ and $\Sigma_2(0)=\Sigma_2^*$. According to Eq. (\ref{Eq37}), we see that $\Sigma_1^*$ is a function of $\Sigma_1(0)$. Thus, by taking advantage of Newton--Raphson method, we can calculate $\Sigma_1(0)$ from an initial guess.  
\item Step II. Performing forward integration under the unperturbed Hamiltonian over one period of $\sigma_1$ (taking arbitrary value for $\sigma_2(0)$ at this step), it is possible to determine the periods ($T_1$ for $\sigma_1$ and $T_2$ for $\sigma_2$). According to the relation $\sigma_1^* = 2\pi/T_1*t$, the time $t$ can be determined.
\item Step III. The relation $\sigma_2^* = \sigma_2^*(0) + 2\pi/T_2*t$ leads to $\sigma^*_2(0)$. Thus $\sigma_2(0)$ is determined. 
\item Step IV. At last, the state $\left(\sigma_1,\sigma_2,\Sigma_1,\Sigma_2\right)$ at the current time $t$ can be obtained by performing forward integration from the initial state $(\sigma_1(0) = 0, \sigma_2(0),\Sigma_1(0),\Sigma_2(0)=\Sigma_2^*)$.
\end{itemize}

\begin{figure*}
\centering
\includegraphics[width=\columnwidth]{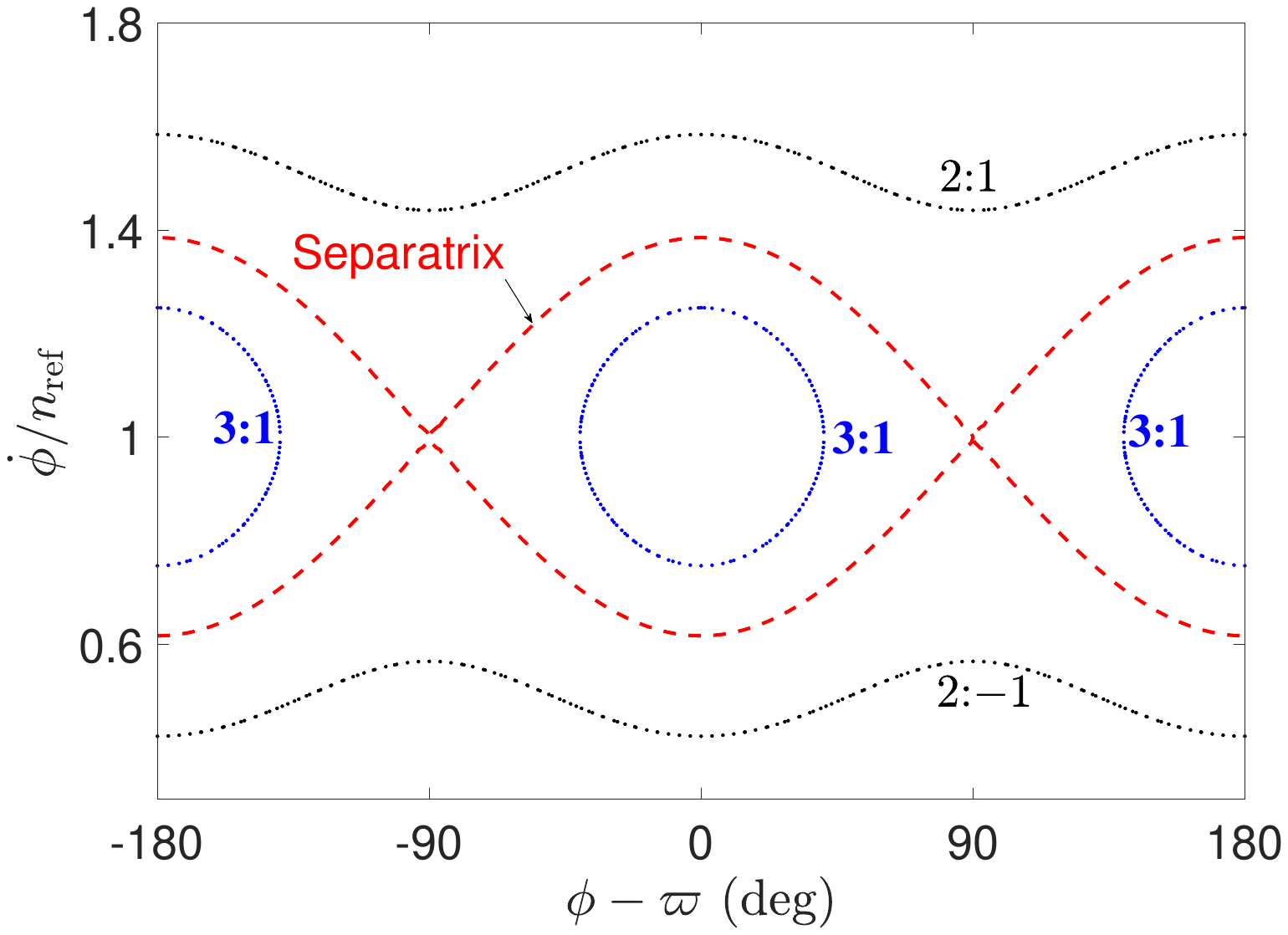}
\includegraphics[width=\columnwidth]{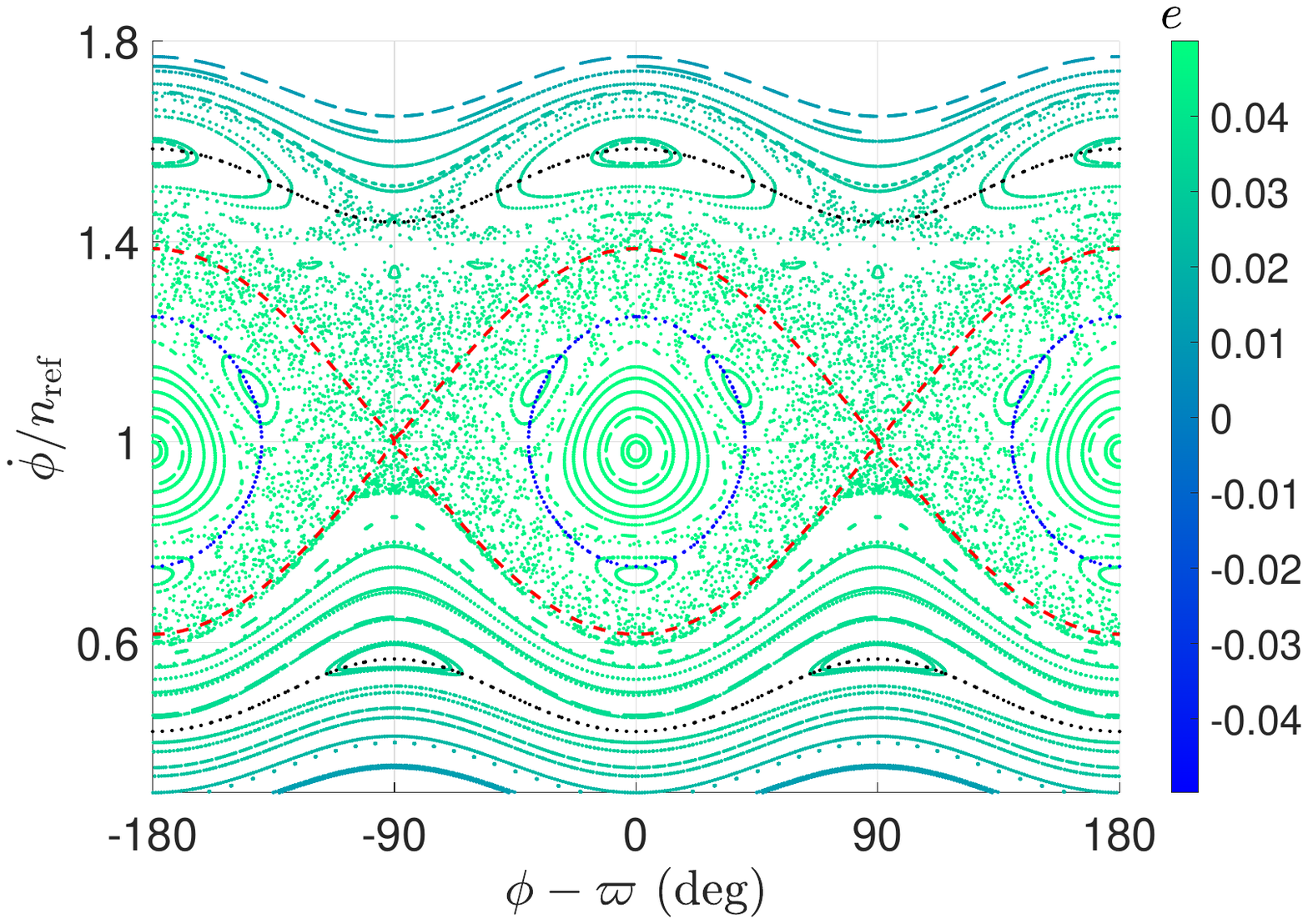}
\caption{Nominal location of high-order and secondary spin-orbit resonances (\textit{left panel}) and comparison to Poincar\'e section (\textit{right panel}). Red dashed lines stand for the dynamical separatrix of the synchronous (1:1) resonance.}
\label{Fig7}
\end{figure*}

Performing the direct transformation,
\begin{equation}\label{Eq39}
\left(\sigma_1,\sigma_2,\Sigma_1,\Sigma_2\right) \Rightarrow \left(\sigma_1^*,\sigma_2^*,\Sigma_1^*,\Sigma_2^*\right), 
\end{equation}
we can write the unperturbed Hamiltonian (\ref{Eq29}) in the normal form,
\begin{equation}\label{Eq40}
{{\cal H}_0}\left( {{\sigma _1},{\Sigma _1},{\Sigma _2}} \right) \Rightarrow {{\cal H}_0}\left( {\Sigma _1^*,\Sigma _2^*} \right)
\end{equation}
where the angular coordinates $\sigma_1^*$ and $\sigma_2^*$ are both cyclic variables. Therefore, the fundamental frequencies can be determined as follows:
\begin{equation}\label{Eq41}
\begin{aligned}
{\dot \sigma} _1^{\rm{*}} &= \frac{{\partial {{\cal H}_0}\left( {\Sigma _1^*,\Sigma _2^*} \right)}}{{\partial \Sigma _1^*}} = \frac{{2\pi }}{{{T_1}}},\\
{\dot \sigma} _2^{\rm{*}} &= \frac{{\partial {{\cal H}_0}\left( {\Sigma _1^*,\Sigma _2^*} \right)}}{{\partial \Sigma _2^*}} = \frac{{2\pi }}{{{T_2}}}.
\end{aligned}
\end{equation}
Furthermore, resonances may happen if the fundamental frequencies satisfy a commensurability relation (resonance condition): 
\begin{equation}\label{Eq42}
{k_1}\dot \sigma _1^{*} - {k_2}\dot \sigma _2^{*} = 0,
\end{equation}
where $k_1 \in \mathbb{N}$ and $k_2 \in \mathbb{Z}$. By solving Eq. (\ref{Eq42}), it is possible to determine the nominal location of $k_1$:$k_2$ spin-orbit resonances between $\sigma_1^*$ and $\sigma_2^*$. The web of resonances are shown in Fig. \ref{Fig7}, where Poincar\'e section is provided as background in the right panel. Dynamical separatrix associated with synchronous spin-orbit resonance is marked in red dashed line. In particular, the resonances occupying inside the region bounded by curves of separatrix are referred to as secondary spin-orbit resonances, and the ones occupying outside the region bounded by curves of separatrix are called high-order spin-orbit resonances. It is observed from Fig. \ref{Fig7} that (a) 3:1 resonance can be found inside the synchronous resonance, and 2:1 and 2:-1 resonances can be found outside the synchronous resonance, (b) distribution of resonance curves is symmetric with respect to the line of $\dot\phi/n_{\rm ref} = 1.0$, and (c) there is a good correspondence between the numerical structures arising in Poincar\'e section and distribution of resonance curves. 

In the coming subsection, we aim to study dynamical structures associated with high-order and/or secondary spin-orbit resonances by taking advantage of perturbative treatments \citep{henrard1990semi}.

\subsection{Resonant Hamiltonian}
\label{Sect4-4}

Performing the canonical transformation defined by Eq. (\ref{Eq37}), we can express the Hamiltonian as
\begin{equation}\label{Eq43}
{\cal H} = {{\cal H}_0}\left( {\Sigma _1^*,\Sigma _2^*} \right) + \varepsilon {{\cal H}_1}\left( {\sigma _1^*,\sigma _2^*,\Sigma _1^*,\Sigma _2^*} \right)
\end{equation}
which is a standard form of Hamiltonian, composed of the unperturbed Hamiltonian ${\cal H}_0 \left( {\Sigma _1^*,\Sigma _2^*} \right)$ and the perturbation part ${{\cal H}_1}\left( {\sigma _1^*,\sigma _2^*,\Sigma _1^*,\Sigma _2^*} \right)$. Thus, it is ready for us to address the problem at hand by means of perturbation theory \citep{henrard1986perturbation,henrard1990semi}. 

In order to study the high-order and/or secondary spin-orbit resonances, let us introduce the following canonical transformations:
\begin{equation}\label{Eq44}
\begin{aligned}
{\gamma _1} &= \sigma _1^{*} - \frac{{{k_2}}}{{{k_1}}}\sigma _2^{*},\quad {\Gamma _1} = \Sigma _1^*,\\
{\gamma _2} &= \sigma _2^{*},\quad {\Gamma _2} = \Sigma _2^* + \frac{{{k_2}}}{{{k_1}}}\Sigma _1^*,
\end{aligned}
\end{equation}
with the generating function as
\begin{equation*}
{\cal S} = \sigma _1^{*}{\Gamma _1} + \sigma _2^{*}\left( {{\Gamma _2} - \frac{{{k_2}}}{{{k_1}}}{\Gamma _1}} \right).
\end{equation*}
Evidently, $\gamma_1$ is the argument of $k_1$:$k_2$ spin-orbit resonance between $\sigma_1^*$ and $\sigma_2^*$. In terms of the set of canonical variables $(\gamma_1,\gamma_2,\Gamma_1,\Gamma_2)$, the Hamiltonian (\ref{Eq43}) can be further organized as follows:
\begin{equation}\label{Eq45}
{\cal H} = {{\cal H}_0}\left( {{\Gamma _1},{\Gamma _2}} \right) + \varepsilon {{\cal H}_1}\left( {{\gamma _1},{\gamma _2},{\Gamma _1},{\Gamma _2}} \right).
\end{equation}
In particular, when a binary asteroid system is located inside the $k_1$:$k_2$ resonance, it holds $\dot \gamma_1 \approx 0$, showing that $\dot \gamma_1 \ll \dot \gamma_2$. As a result, Eq. (\ref{Eq45}) determines a separable  Hamiltonian model \citep{henrard1990semi}, where $(\sigma_1,\Sigma_1)$ stands for the slow degree of freedom and $(\sigma_2,\Sigma_2)$ represents the fast degree of freedom.  

Regarding a separable Hamiltonian model, resonant Hamiltonian can be formulated by taking advantage of averaging theory \citep{henrard1990semi},
\begin{equation}\label{Eq46}
\begin{aligned}
{{\cal H}^*}\left( {{\gamma _1},{\Gamma _1},{\Gamma _2}} \right) =& {{\cal H}_0}\left( {{\Gamma _1},{\Gamma _2}} \right)\\
&+ \frac{1}{{2{k_1}\pi }}\int\limits_0^{2{k_1}\pi } {{\varepsilon{\cal H}_1}\left( {{\gamma _1},{\gamma _2},{\Gamma _1},{\Gamma _2}} \right){\rm d}{\gamma _2}}.
\end{aligned}
\end{equation}
Under such a resonant Hamiltonian model, the angular coordinate $\gamma_2$ is a cyclic variable, showing that its conjugate momentum
\begin{equation}\label{Eq47}
{\Gamma _2} = \Sigma _2^* + \frac{{{k_2}}}{{{k_1}}}\Sigma _1^*
\end{equation}
is a motion integral. As a result, Eq. (\ref{Eq46}) determines an integrable Hamiltonian model, depending on the motion integral ${\Gamma _2}$.

\subsection{Phase-space structures}
\label{Sect4-5}

The resonant Hamiltonian (\ref{Eq46}) determines an integrable model. Thus, it is possible to explore dynamical structures by means of drawing phase portraits. In practice, phase portraits are produced by plotting level curves of resonant Hamiltonian $\cal H^*$ in the $(\gamma_1,\Gamma_1)$ space with given motion integral $\Gamma_2$.

\begin{figure*}
\centering
\includegraphics[width=\columnwidth]{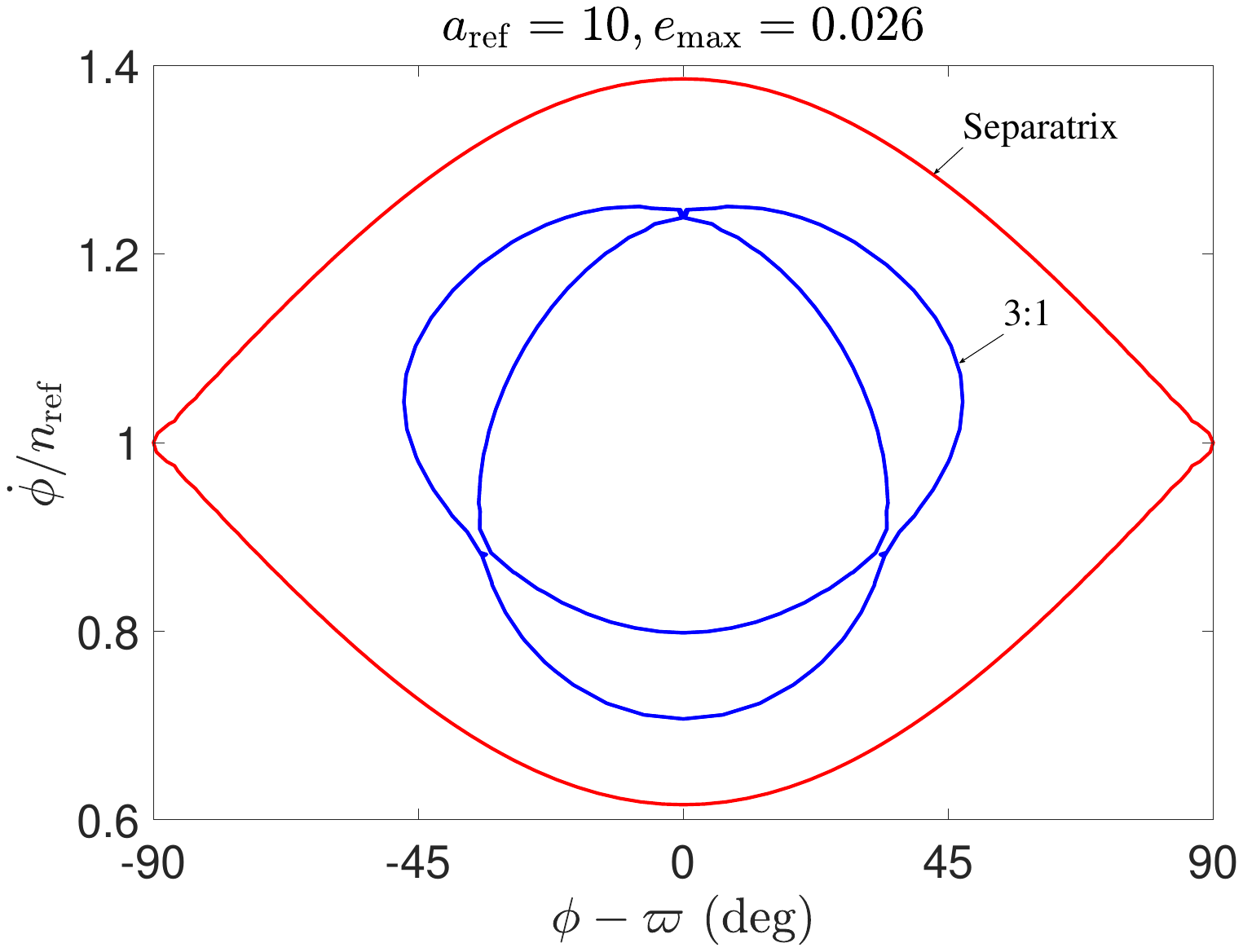}
\includegraphics[width=\columnwidth]{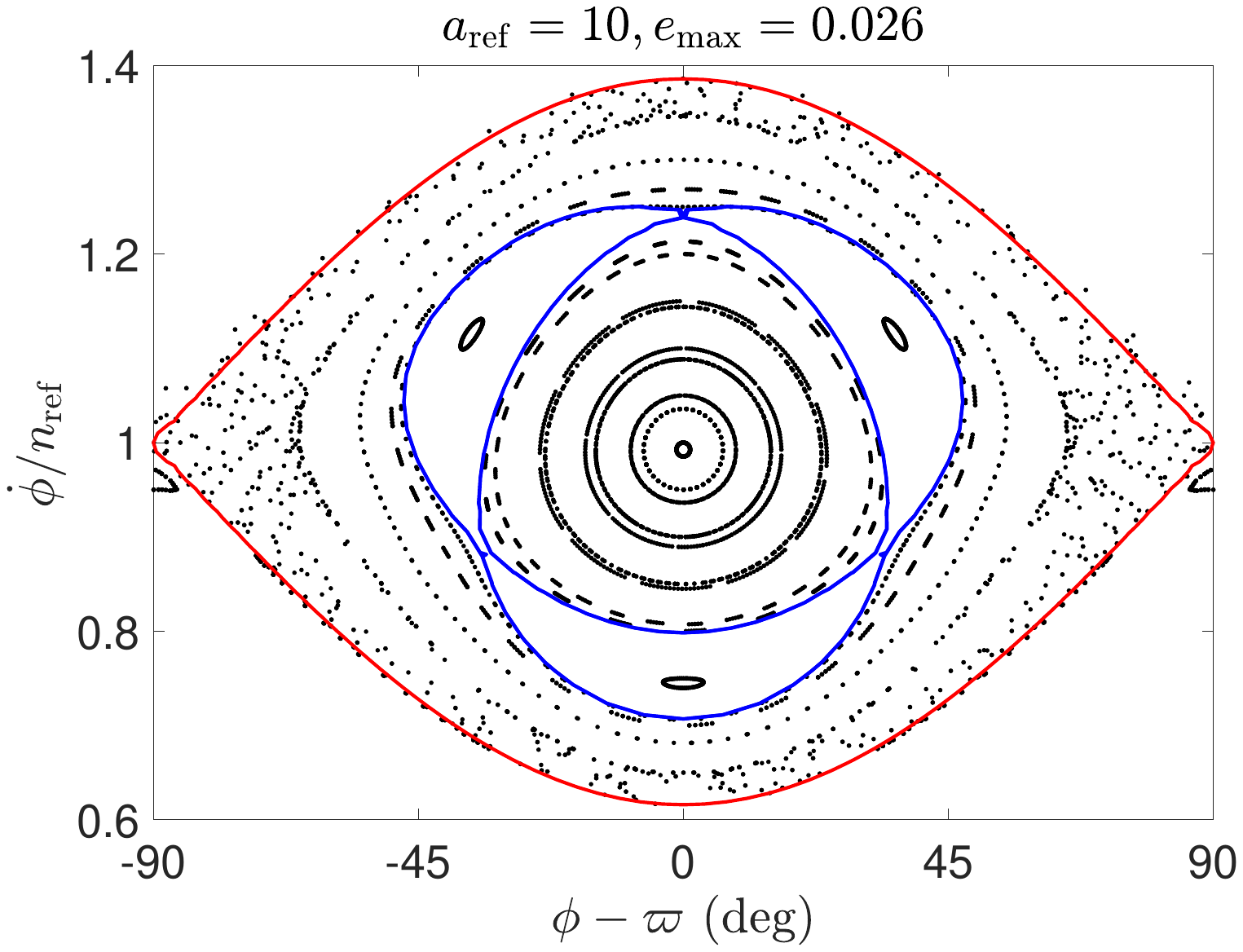}\\
\includegraphics[width=\columnwidth]{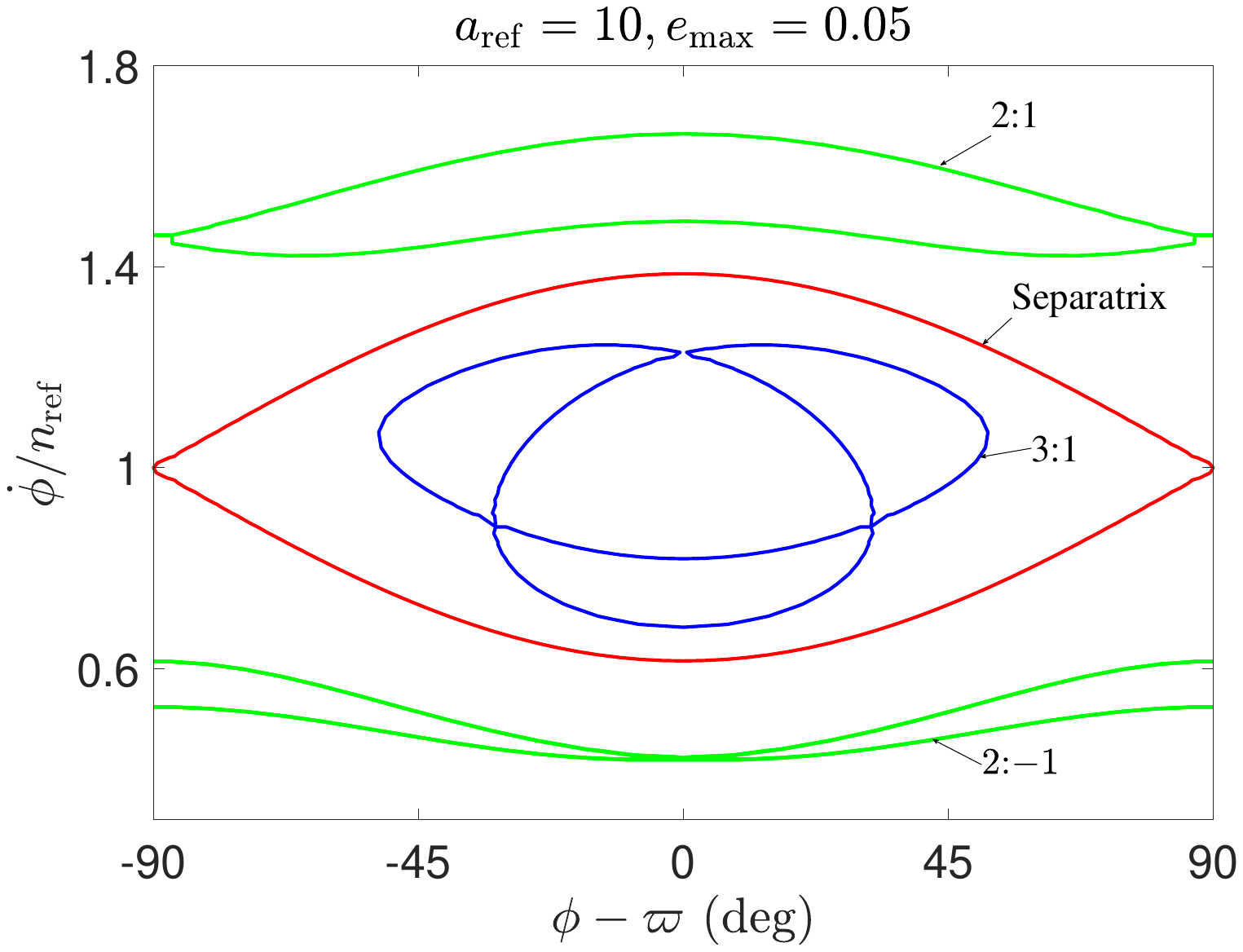}
\includegraphics[width=\columnwidth]{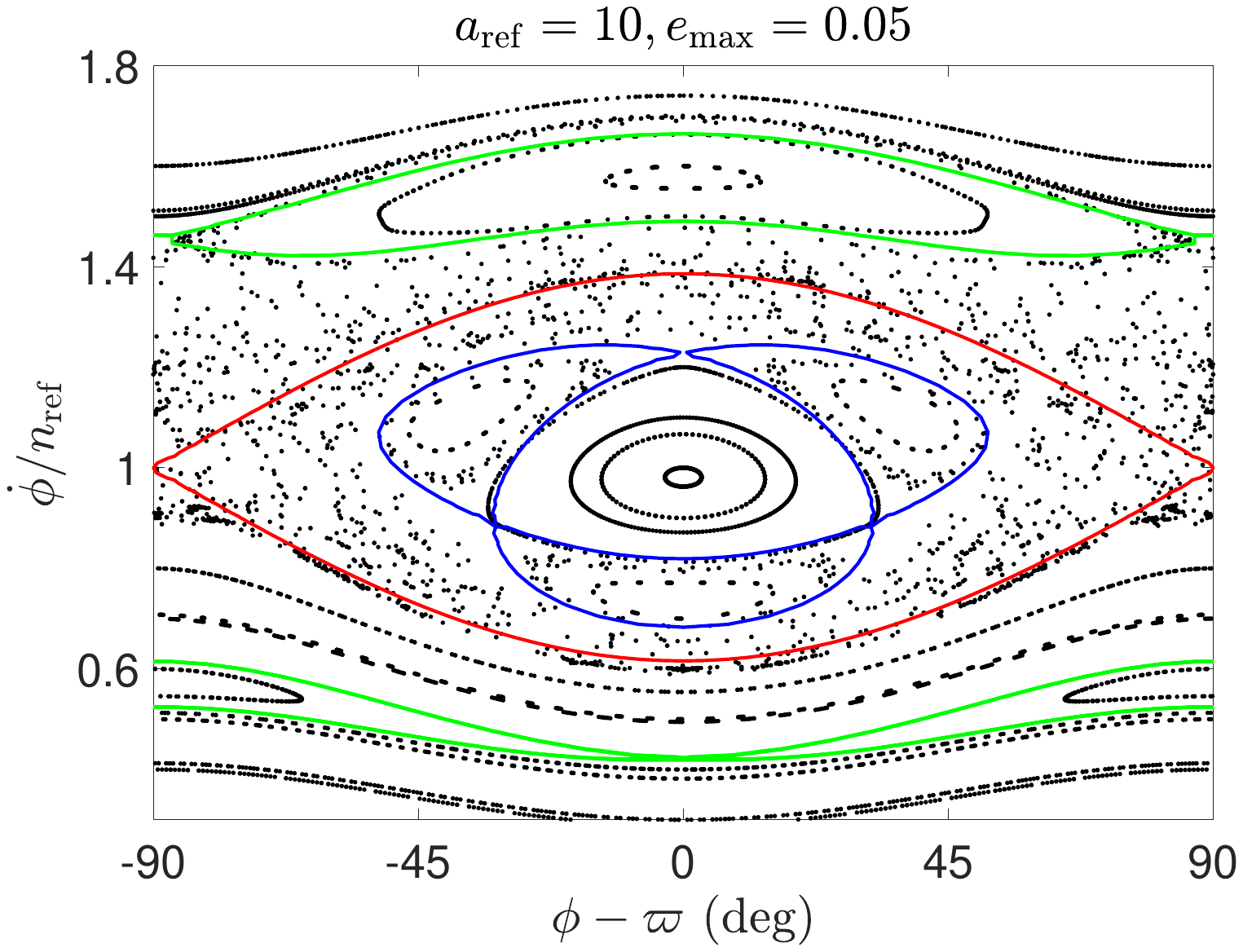}
\caption{Analytical structures arising in phase portraits (\textit{left-column panels}) and comparison to numerical structures arising in Poincar\'e sections (\textit{right-column panels}) at different maximum eccentricities when the total angular momentum is characterized by $a_{\rm ref} = 10$. Red dashed lines stand for the separatrices of the synchronous (1:1) resonance, blue lines stand for separatrices of secondary spin-orbit resonances and green lines represent separatrices of high-order spin-orbit resonances. An excellent agreement can be observed between analytical and numerical structures.}
\label{Fig8}
\end{figure*}

For convenience, we need to project the phase space $(\gamma_1,\Gamma_1)$ to the $(\phi-\varpi,\dot\phi)$ space by considering the definition of Poincaré sections (see Eq. \ref{Eq17}). Results are presented in Fig. \ref{Fig8}, where red lines stand for the dynamical separatrices of synchronous spin-orbit resonance (see Eq. \ref{Eq32-1}), blue lines stand for the separatrices of secondary 3:1 resonance and green lines stand for separatrices of high-order 2:1 and 2:-1 resonances. In particular, the total angular momentum is specified by $a_{\rm ref} = 10$. Two levels of Hamiltonian specified by $e_{\max}=0.026$ and $e_{\max}=0.05$ are taken into account. For convenience of comparison, the associated Poincar\'e sections are provided in the right panel as background.

From Fig. \ref{Fig8}, it is observed that (a) there is an excellent agreement between analytical structures arising in phase portraits and numerical structures arising in Poincar\'e sections, meaning that the numerical structures can be well analytically reproduced, and (b) the islands of libration associated with high-order or secondary spin-orbit resonances are well bounded by the analytical separatrices. 

As a result, the resonant Hamiltonian provides us a reliable tool to analytically understand and predict dynamical structures in phase space. In the coming section, analytical solutions are applied to three binary asteroid systems.

\begin{table*}
\centering
\caption{Physical parameters (in normalized units) of three binary asteroid systems adopted in this work. Please refer to http://www.asu.cas.cz/$\sim$asteroid/binastdata.htm for detailed data of binary asteroid systems.}
\begin{tabular*}{\textwidth}{@{\extracolsep{\fill}}lcccccc@{\extracolsep{\fill}}}
\hline
{binary system}&{$a_s$}&{$b_s$}&{$\alpha_m$}&{$a_{\rm ref}$}&{$\Delta \dot\phi/n_{\rm ref}$(or $\alpha$)}&{$L_{\rm rot}/L_{\rm orb}$}\\
\hline
(65803) Didymos&1.0&0.6452&0.0116&13.4118&1.1&$2.2 \times 10^{-3}$\\
(80218) ${\rm VO}_{123}$&1.0&0.6579&0.0032&19.2857&1.0897&$1.1 \times 10^{-3}$\\
(4383) Suruga&1.0&0.4902&0.0066&20.0&1.36& $1.0 \times 10^{-3}$\\
\hline
\end{tabular*}
\label{Tab1}
\end{table*}

\section{Applications}
\label{Sect5}

Analytical study discussed in Sect. \ref{Sect4} is applied to three example binary asteroid systems, including (65803) Didymos, (80218) ${\rm VO}_{123}$ and (4383) Suruga \citep{warner2013something,pravec2016binary,pravec2019asteroid}. Please refer to Table \ref{Tab1} for their parameters, including the major axes $(a_s,b_s)$ in normalized units, mass ratio $\alpha_m$, the reference semimajor axis $a_{\rm ref}$, the resonant half-width of synchronous spin-orbit resonance $\Delta \dot\phi/n_{\rm ref}$ (or asphericity parameter $\alpha$) and ratio of rotation-to-orbit angular momentum $L_{\rm rot}/L_{\rm orb}$. It shows that, among these three systems, Suruga holds the largest equatorial elongation, meaning that it has the strongest strength of synchronous spin-orbit resonance. 

For these systems, both the minor axis $c_s$ of the ellipsoidal secondary and the orbital eccentricity $e$ are not determinant, thus they are not presented in Table \ref{Tab1}. In practical simulations, we take $c_s = b_s$. Considering the fact that the orbital eccentricity of the secondary is usually small and highly uncertain, thus in practical simulations it is controlled by the maximum eccentricity $e_{\max}$. In particular, two cases of $e_{\max}=0.04,0.05$ are taken for Dydymos, two cases of $e_{\max}=0.03,0.04$ are taken for both ${\rm VO}_{123}$ and Suruga. With a given pair of $(a_{\rm ref}, e_{\max})$, both the total angular momentum and the Hamiltonian are determinant. According to the discussions made in Sect. \ref{Sect3}, it is known that a higher value of $e_{\max}$ indicates a larger area of physically allowed region.

\begin{figure*}
\centering
\includegraphics[width=\columnwidth]{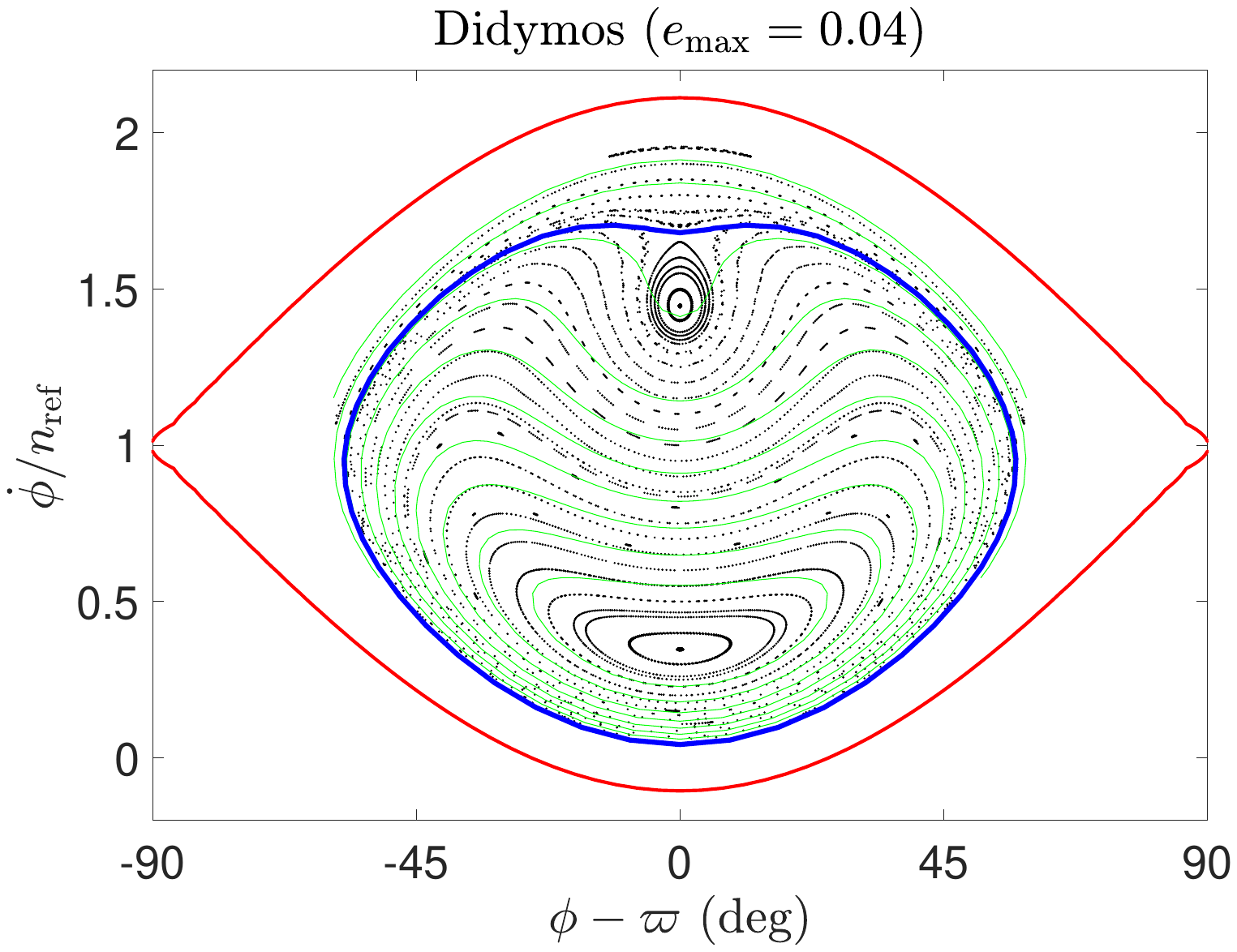}
\includegraphics[width=\columnwidth]{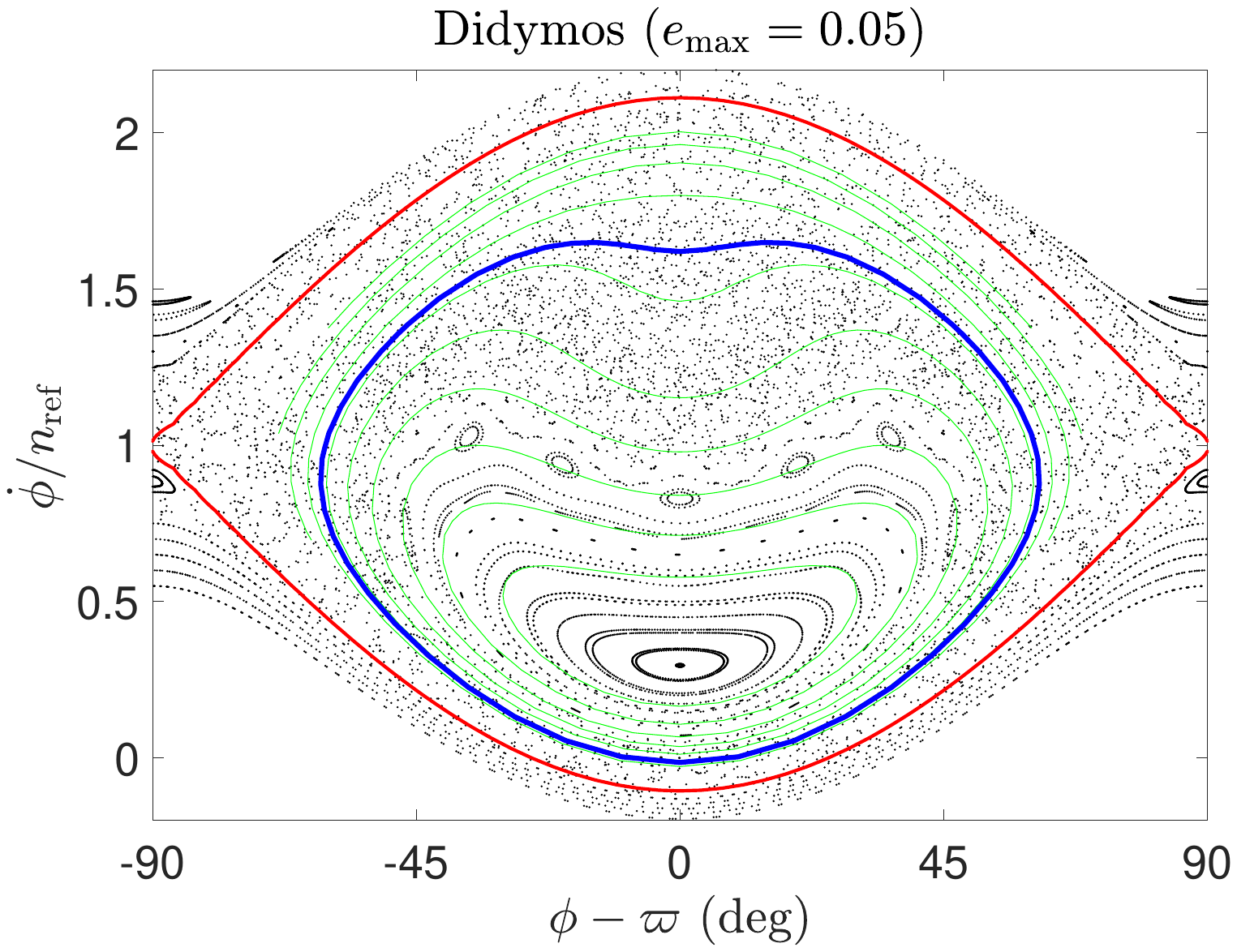}\\
\includegraphics[width=\columnwidth]{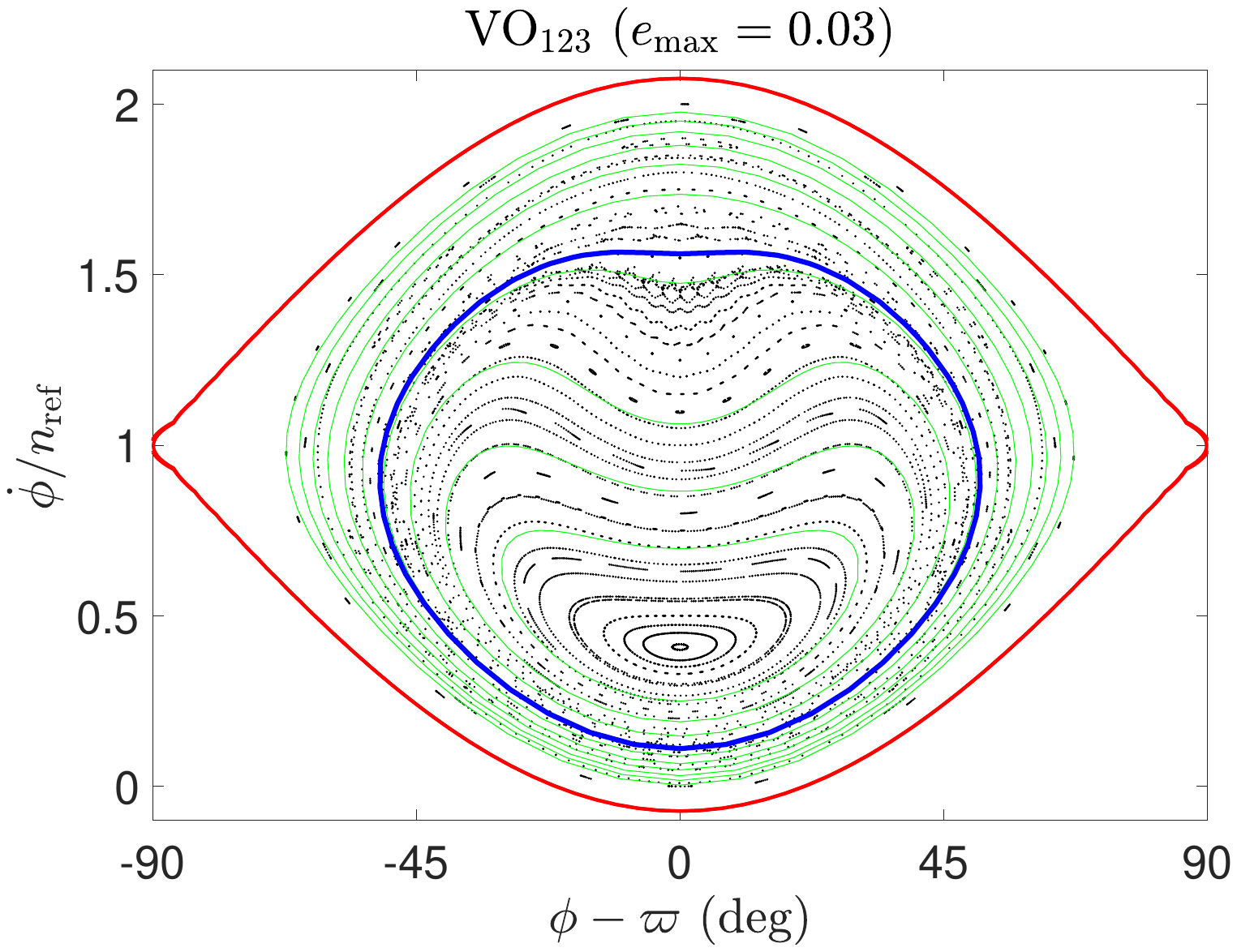}
\includegraphics[width=\columnwidth]{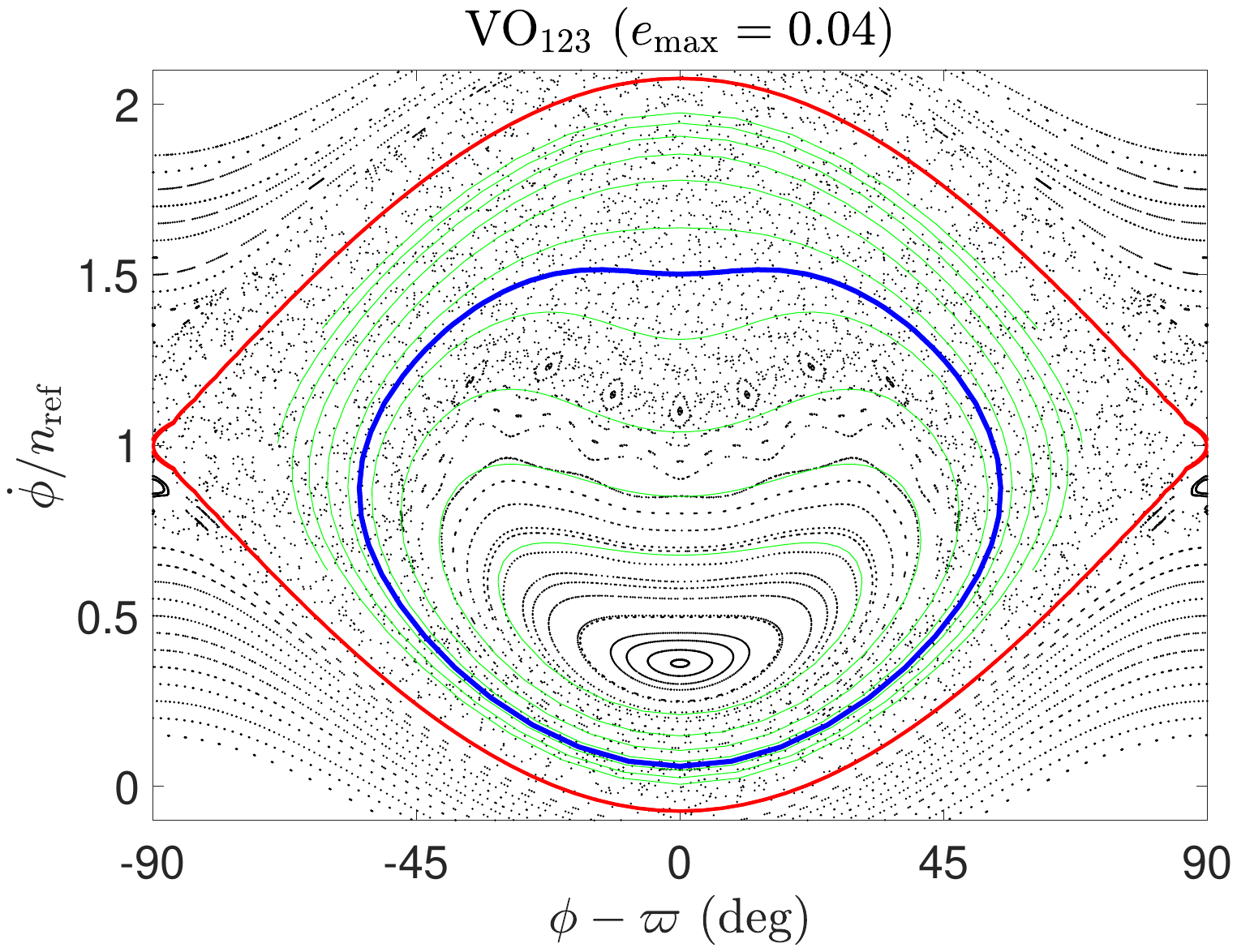}\\
\includegraphics[width=\columnwidth]{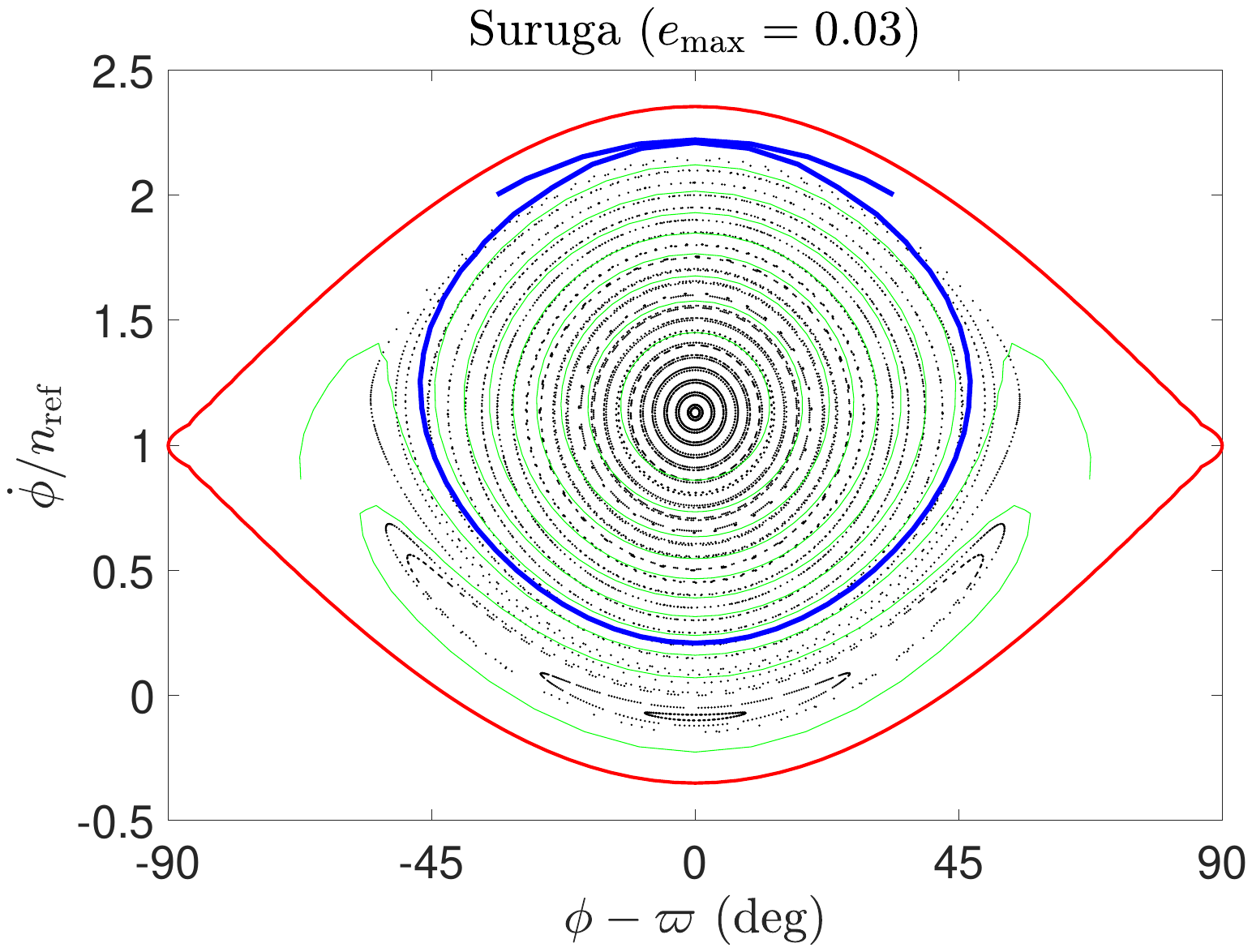}
\includegraphics[width=\columnwidth]{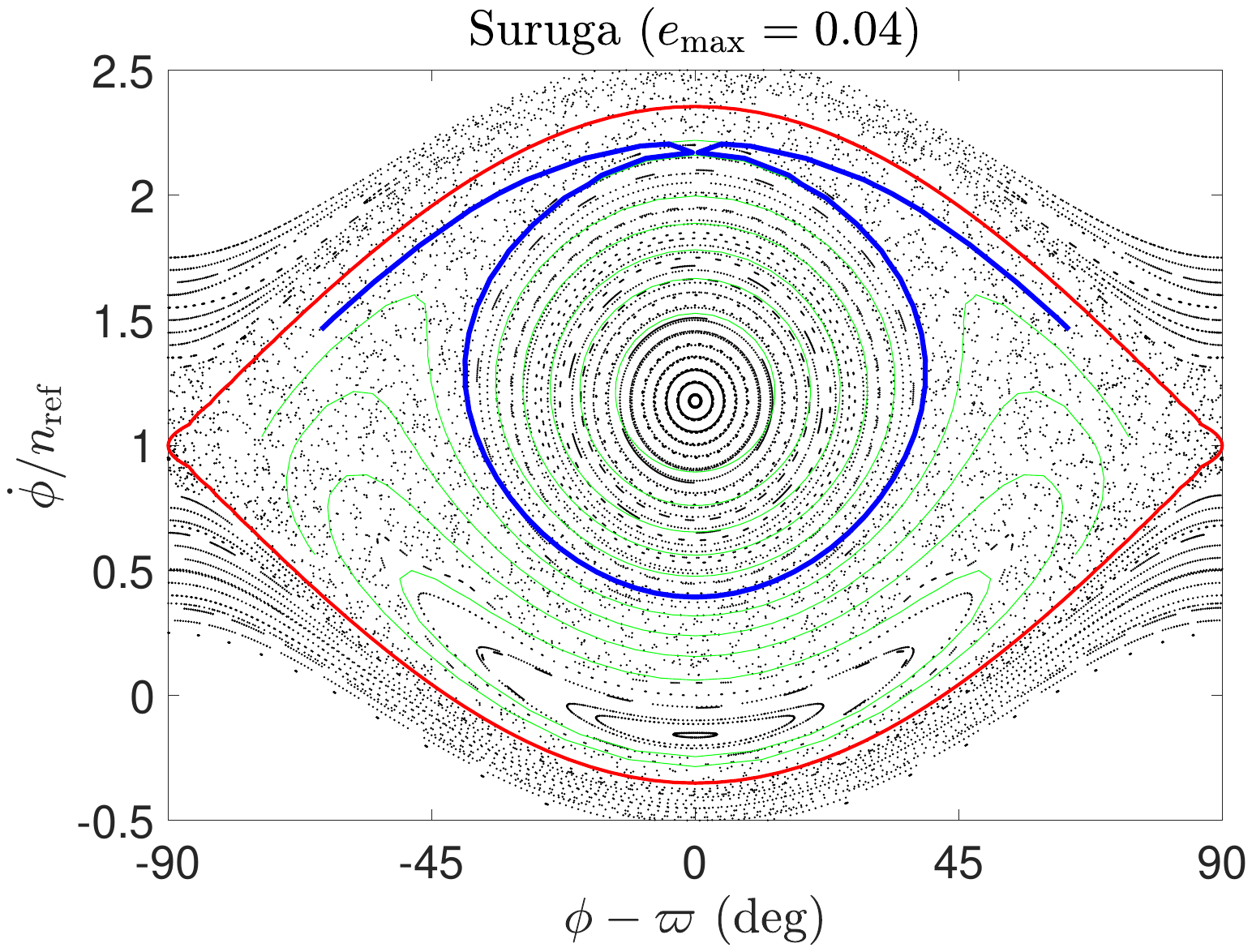}
\caption{Analytical structures arising in phase portraits and numerical structures arising in Poincar\'e sections for three representative binary asteroid systems (Didymos, ${\rm VO}_{123}$ and Suruga). Red dashed lines stand for the dynamical separatrices of the synchronous (1:1) resonance and the blue lines represent the separatrices of the secondary resonances. Green lines are level curves of resonant Hamiltonian (i.e., phase portraits). Chaotic motions can be observed in the regions around dynamical separatrices of the primary synchronous (1:1) resonance as well as around dynamical separatrices of secondary resonance.}
\label{Fig9}
\end{figure*}

Simulation results are presented in Fig. \ref{Fig9}, including the phase portraits (see the green lines) as well as the associated Poincar\'e sections (see the black dots). In addition, the dynamical separatrices of the synchronous spon-orbit resonance are marked by red lines, and the dynamical separatrices of secondary 1:1 spin-orbit resonance are marked by blue lines. It is observed that (a) secondary 1:1 spin-orbit resonance can be found for three binary asteroid systems, showing that the secondary 1:1 resonance is responsible for the structures inside synchronous spin-orbit resonance, (b) there is a perfect agreement between analytical structures arising in phase portraits and numerical structures arising in Poincar\'e sections, and (c) chaotic motions are distributed around dynamical separatrices of the primary and/or secondary spin-orbit resonance and the area of chaotic region is larger with a higher value of $e_{\max}$. 

Recently, \citet{lei2024dynamical} studied dynamical structures associated with high-order and/or secondary spin-orbit resonances for binary asteroid systems Didymos and Suruga under the restricted configuration (i.e., the orbit of the secondary remains invariant). Comparing to their results (see Fig. 8 in their work), we can see that the dynamical structures are qualitatively similar. This is because the orbital angular momentum is much larger than the rotational angular momentum about these two example systems, leading to a relatively weak coupling between translation and rotation, as discussed in \citet{lei2024dynamical}. The dynamical closeness of asteroid pairs is discussed in \citet{jafari2023surfing} by introducing a criterion: a pair of asteroids is dynamically close when the
ratio of rotational angular momentum to the orbital angular momentum is greater than 10 percent. Based on the physical data provided in Table \ref{Tab1}, we can get the rotational-to-translational ratio of angular momentum is much smaller than 10 percent, indicating that the three binary asteroid systems are indeed not dynamically close. As expected, the problem of spin-orbit resonance is a good approximation for them.

Even through the coupling is weak, we can also see the following differences between the problems of spin-orbit coupling (non-restricted configuration) and the spin-orbit resonance (restricted configuration): (a) it is a single degree-of-freedom non-autonomous Hamiltonian system for the problem of spin-orbit resonance (or equivalently it is a 1.5 degree-of-freedom autonomous Hamiltonian model because the time stands for an additional 0.5 degree of freedom), while it is a 2 degree-of-freedom autonomous Hamiltonian system depending on the total angular momentum for the problem of spin-orbit coupling; (b) there is no physically forbidden region for the problem of spin-orbit resonance, while the physically allowed region is determined by the total angular momentum and Hamiltonian specified by $(a_{\rm ref},e_{\max})$ for the spin-orbit coupling problem; (c) in the spin-orbit resonance problem the orbital semimajor axis $a$ and eccentricity $e$ of the secondary remain constant, but they are changed under the spin-orbit coupling problem in order to make the total angular momentum and Hamiltonian be conserved.

\section{Conclusions}
\label{Sect6}

In this work, dynamics of spin-obit coupling is thoroughly studied in binary asteroid systems consisting of spherical primaries and ellipsoidal secondaries. Under the planar assumption, the Hamiltonian of system is formulated up to the fourth degree and order by taking advantage of elliptic expansions. The resulting Hamiltonian for describing spin-orbit coupling determines a two-degree-of-freedom dynamical model, depending on the total angular momentum. Conservation of the total angular momentum implies that the orbital and rotational angular momentum can exchange with each other, showing that the semimajor axis and eccentricity of mutual orbit are no longer invariant. The strength of spin-orbit coupling in a binary asteroid system is determined by the dynamical closeness of asteroid pairs, which can be measured by means of the ratio of rotational-to-orbital angular momentum \citep{jafari2023surfing}. 

A pair of parameters $(a_{\rm ref}, e_{\max})$ is introduced to characterize the total angular momentum $G_{\rm tot}$ and the magnitude of Hamiltonian ${\cal H}$. The total angular momentum and Hamiltonian, specified by $(a_{\rm ref}, e_{\max})$, together determine the physically allowed region of motion. In particular, when the total angular momentum is given (i.e., $a_{\rm ref}$ is provided), the feasible region increases with $e_{\max}$. With a given pair of $(a_{\rm ref}, e_{\max})$, it is found that the maximum eccentricity happens at the libration center of synchronous spin-orbit resonance and the minimum eccentricity (equal to zero) takes place at the borders of physically allowed region.

The technique of Poincar\'e section is taken to explore dynamical structures of spin-orbit coupling problem. In particular, the influences of four parameters upon dynamical structures are discussed, including the reference semimajor axis $a_{\rm ref}$, the maximum eccentricity $e_{\max}$, the secondary-to-primary mass ratio $\alpha_m$ and the shape parameter $b_s$. It is found that the parameters $a_{\rm ref}$ and $e_{\max}$ influence the dynamical structures of Poincar\'e section through controlling the physically allowed region of motion, while the system parameters $\alpha_m$ and $b_s$ influence the types of libration islands associated with high-order and/or secondary spin-orbit resonances.

To understand dynamical structures arising in Poincar\'e sections, analytical study is performed by taking advantage of perturbative treatments. In particular, the Hamiltonian terms associated with the synchronous spin-orbit resonance are included in the unperturbed Hamiltonian. A series of canonical transformations are performed and, at last, the resonant Hamiltonian is formulated based on averaging theory. The resonant Hamiltonian determines an integrable dynamical model, where the phase-space structures can be revealed by producing phase portraits, i.e., plotting level curves of resonant Hamiltonian with a given motion integral. Phase portraits can help to understand the emergence and dynamical effects of high-order and/or secondary spin-orbit resonances. Simulation results show that there is a perfect agreement between analytical structures arising in phase portraits and numerical structures arising in Poincar\'e sections.

Analytical solutions are applied to three binary asteroid systems, including (65803) Didymos, (80218) ${\rm VO}_{123}$ and (4383) Suruga. The secondaries of these three examples hold high equatorial elongation (i.e., asphericity parameter $\alpha$ is high), showing that their synchronous spin-orbit resonances have large libration zone. Analytical and numerical results show that there is a high possibility for them to locate inside secondary 1:1 spin-orbit resonance. For these binary asteroid systems, their ratios of rotation-to-orbit angular momentum are very small, showing that the spin-orbit coupling is weak. As expected, the problem of spin-orbit resonance may provide reliable approximation for them, as explored in \citet{lei2024dynamical}.

%\begin{acknowledgments}
%The author wishes to thank Prof. Xiyun Hou for helpful discussions in preparing the manuscript. This work is supported by the %National Natural Science Foundation of China (Nos. 12073011 and 12233003) and the National Key R\&D Program of China (No. %2019YFA0706601).
%\end{acknowledgments}

%Appendices can be broken into separate sections just like in the main text.
%The only difference is that each appendix section is indexed by a letter
%(A, B, C, etc.) instead of a number.  Likewise numbered equations have
%the section letter appended.  Here is an equation as an example.
%\begin{equation}
%I = \frac{1}{1 + d_{1}^{P (1 + d_{2} )}}
%\end{equation}
%Appendix tables and figures should not be numbered like equations. Instead
%they should continue the sequence from the main article body.


\bibliography{mybib}{}

\begin{thebibliography}{}
\expandafter\ifx\csname natexlab\endcsname\relax\def\natexlab#1{#1}\fi
\providecommand{\url}[1]{\href{#1}{#1}}
\providecommand{\dodoi}[1]{doi:~\href{http://doi.org/#1}{\nolinkurl{#1}}}
\providecommand{\doeprint}[1]{\href{http://ascl.net/#1}{\nolinkurl{http://ascl.net/#1}}}
\providecommand{\doarXiv}[1]{\href{https://arxiv.org/abs/#1}{\nolinkurl{https://arxiv.org/abs/#1}}}

\bibitem[{Balmino(1994)}]{balmino1994gravitational}
Balmino, G. 1994, CeMDA, 60, 331

\bibitem[{Batygin \& Morbidelli(2015)}]{batygin2015spin}
Batygin, K., \& Morbidelli, A. 2015, ApJ, 810, 110

\bibitem[{Celletti(1990{\natexlab{a}})}]{Celletti1990AnalysisI}
Celletti, A. 1990{\natexlab{a}}, Z Ang Math Phys, 41, 174

\bibitem[{Celletti(1990{\natexlab{b}})}]{Celletti1990AnalysisII}
---. 1990{\natexlab{b}}, Z Ang Math Phys, 41, 453

\bibitem[{Celletti \& Chierchia(2000)}]{celletti2000hamiltonian}
Celletti, A., \& Chierchia, L. 2000, CeMDA, 76, 229

\bibitem[{Chirikov(1979)}]{chirikov1979universal}
Chirikov, B.~V. 1979, PR, 52, 263

\bibitem[{Correia {et~al.}(2015)Correia, Leleu, Rambaux, \&
  Robutel}]{correia2015spin}
Correia, A.~C., Leleu, A., Rambaux, N., \& Robutel, P. 2015, A\&A, 580, L14

\bibitem[{{\'C}uk \& Burns(2005)}]{cuk2005effects}
{\'C}uk, M., \& Burns, J.~A. 2005, Icar, 176, 418

\bibitem[{Fang \& Margot(2011)}]{fang2011near}
Fang, J., \& Margot, J.-L. 2011, AJ, 143, 24

\bibitem[{Flynn \& Saha(2005)}]{flynn2005second}
Flynn, A.~E., \& Saha, P. 2005, AJ, 130, 295

\bibitem[{Gkolias {et~al.}(2016)Gkolias, Celletti, Efthymiopoulos, \&
  Pucacco}]{gkolias2016theory}
Gkolias, I., Celletti, A., Efthymiopoulos, C., \& Pucacco, G. 2016, MNRAS, 459,
  1327

\bibitem[{Gkolias {et~al.}(2019)Gkolias, Efthymiopoulos, Celletti, \&
  Pucacco}]{gkolias2019accurate}
Gkolias, I., Efthymiopoulos, C., Celletti, A., \& Pucacco, G. 2019, CNSNS, 77,
  181

\bibitem[{Goldreich \& Peale(1966)}]{Goldreich1966Spin}
Goldreich, P., \& Peale, S. 1966, AJ, 71, 425

\bibitem[{Henrard(1990)}]{henrard1990semi}
Henrard, J. 1990, CeMDA, 49, 43

\bibitem[{Henrard \& Lemaitre(1986)}]{henrard1986perturbation}
Henrard, J., \& Lemaitre, A. 1986, CeMec, 39, 213

\bibitem[{Hou {et~al.}(2017)Hou, Scheeres, \& Xin}]{hou2017mutual}
Hou, X., Scheeres, D.~J., \& Xin, X. 2017, CeMDA, 127, 369

\bibitem[{Hou \& Xin(2017)}]{hou2017note}
Hou, X., \& Xin, X. 2017, AJ, 154, 257

\bibitem[{Jafari-Nadoushan(2023)}]{jafari2023surfing}
Jafari-Nadoushan, M. 2023, MNRAS, 520, 3514

\bibitem[{Jafari-Nadoushan \& Assadian(2015)}]{jafari2015widespread}
Jafari-Nadoushan, M., \& Assadian, N. 2015, NonLD, 81, 2031

\bibitem[{Jafari-Nadoushan \&
  Assadian(2016{\natexlab{a}})}]{nadoushan2016geography}
---. 2016{\natexlab{a}}, Icar, 265, 175

\bibitem[{Jafari-Nadoushan \&
  Assadian(2016{\natexlab{b}})}]{jafari2016chirikov}
---. 2016{\natexlab{b}}, NonLD, 85, 1837

\bibitem[{Kaula(1961)}]{kaula1961analysis}
Kaula, W.~M. 1961, Geophys J Int, 5, 104

\bibitem[{Lei(2024)}]{lei2024dynamical}
Lei, H. 2024, AJ, 167, 121

\bibitem[{Lemaitre {et~al.}(2006)Lemaitre, D’Hoedt, \&
  Rambaux}]{lemaitre20063}
Lemaitre, A., D’Hoedt, S., \& Rambaux, N. 2006, CeMDA, 95, 213

\bibitem[{McMahon \& Scheeres(2010)}]{mcmahon2010secular}
McMahon, J., \& Scheeres, D. 2010, CeMDA, 106, 261

\bibitem[{Morbidelli(2002)}]{morbidelli2002modern}
Morbidelli, A. 2002, Modern celestial mechanics: aspects of solar system
  dynamics (Taylor \& Francis, London and New York)

\bibitem[{Murray \& Dermott(1999)}]{murray1999solar}
Murray, C.~D., \& Dermott, S.~F. 1999, Solar system dynamics (Cambridge
  university press)

\bibitem[{Naidu \& Margot(2015)}]{naidu2015near}
Naidu, S.~P., \& Margot, J.-L. 2015, AJ, 149, 80

\bibitem[{Peale \& Gold(1965)}]{peale1965rotation}
Peale, S., \& Gold, T. 1965, Natur, 206, 1240

\bibitem[{Peale(1969)}]{peale1969generalized}
Peale, S.~J. 1969, AJ, 74, 483

\bibitem[{Peale(1977)}]{peale1977rotation}
Peale, S.~J. 1977, in IAU colloq. 28: Planetary satellites, 87

\bibitem[{Pravec {et~al.}(2016)Pravec, Scheirich, Ku{\v{s}}nir{\'a}k, Hornoch,
  Gal{\'a}d, Naidu, Pray, Vil{\'a}gi, Gajdo{\v{s}}, Korno{\v{s}},
  {et~al.}}]{pravec2016binary}
Pravec, P., Scheirich, P., Ku{\v{s}}nir{\'a}k, P., {et~al.} 2016, Icar, 267,
  267

\bibitem[{Pravec {et~al.}(2019)Pravec, Fatka, Vokrouhlick{\`y}, Scheirich,
  {\v{D}}urech, Scheeres, Ku{\v{s}}nir{\'a}k, Hornoch, Gal{\'a}d, Pray,
  {et~al.}}]{pravec2019asteroid}
Pravec, P., Fatka, P., Vokrouhlick{\`y}, D., {et~al.} 2019, Icar, 333, 429

\bibitem[{Scheeres(2007)}]{scheeres2007dynamical}
Scheeres, D.~J. 2007, Icar, 188, 430

\bibitem[{Scheeres {et~al.}(2006)Scheeres, Fahnestock, Ostro, Margot, Benner,
  Broschart, Bellerose, Giorgini, Nolan, Magri,
  {et~al.}}]{scheeres2006dynamical}
Scheeres, D.~J., Fahnestock, E.~G., Ostro, S.~J., {et~al.} 2006, Sci, 314, 1280

\bibitem[{Showalter \& Hamilton(2015)}]{showalter2015resonant}
Showalter, M., \& Hamilton, D. 2015, Natur, 522, 45

\bibitem[{Tan {et~al.}(2023)Tan, Wang, \& Hou}]{tan2023attitude}
Tan, P., Wang, H., \& Hou, X. 2023, Icar, 390, 115289

\bibitem[{Wang \& Hou(2020)}]{wang2020secondary}
Wang, H., \& Hou, X. 2020, MNRAS, 493, 171

\bibitem[{Wang {et~al.}(2022)Wang, Xin, Hou, \& Feng}]{wang2022stability}
Wang, H., Xin, X., Hou, X., \& Feng, J. 2022, CNSNS, 114, 106638

\bibitem[{Warner(2013)}]{warner2013something}
Warner, B.~D. 2013, The Minor planet bulletin, 40, 119

\bibitem[{Wisdom(2004)}]{wisdom2004spin}
Wisdom, J. 2004, AJ, 128, 484

\bibitem[{Wisdom {et~al.}(1984)Wisdom, Peale, \& Mignard}]{wisdom1984chaotic}
Wisdom, J., Peale, S.~J., \& Mignard, F. 1984, Icar, 58, 137

\end{thebibliography}
\bibliographystyle{aasjournal}



\end{document}